\documentclass[a4paper,12pt]{JHEP3} % 10pt is ignored!

\JHEPspecialurl{http://jhep.sissa.it/JOURNAL/JHEP3.tar.gz}

\usepackage{epsfig,multicol,bbm,amsmath,axodraw4j,times,graphicx,amssymb}
\usepackage{dcolumn}% Align table columns on decimal point
\usepackage{bm}% bold math
\usepackage{cite}
\def\cA{\mathcal{A}}
\def\cM{\mathcal{M}}

\def\ZZ{\mathbb{Z}}

\def\stamp{--- {\bf \today} --- {\bf \jobname.tex}}

\def\cSS#1#2{{\mathcal{S}}_{{#1},{#2}}}

\def\BE{\begin{equation}}
\def\EE{\end{equation}}
\def\spa#1.#2{\left\langle#1\,#2\right\rangle}
\def\spb#1.#2{\left[#1\,#2\right]}
\def\spba#1.#2.#3{\left[#1|#2|#3\right\rangle}
\def\spab#1.#2.#3{\left\langle#1|#2|#3\right]}
\def\spaa#1.#2.#3{\left\langle#1|#2|#3\right\rangle}
\def\spbb#1.#2.#3{\left[#1|#2|#3\right]}
\def\lor#1.#2{\left(#1\,#2\right)}

\def\Year{\expandafter\eatPrefix\the\year}
\newcount\hours \newcount\minutes
\def\monthname{\ifcase\month\or
January\or February\or March\or April\or May\or June\or July\or
August\or September\or October\or November\or December\fi}
\def\shortmonthname{\ifcase\month\orx
Jan\or Feb\or Mar\or Apr\or May\or Jun\or Jul\or
Aug\or Sep\or Oct\or Nov\or Dec\fi}

\def\TimeStamp{\hours\the\time\divide\hours by60%
\minutes -\the\time\divide\minutes by60\multiply\minutes by60%
\advance\minutes by\the\time%
${\rm \shortmonthname}\cdot   \if\day<10{}0\fi\the\day\cdot   \the\year
\qquad\the\hours:\if\minutes<10{}0\fi\the\minutes$}

\newskip\humongous \humongous=0pt plus 1000pt minus 100pt

\newif\ifdtup

\newcounter{eqnumber}[section]

\newbox\charbox
\newbox\slabox

\def\spa#1.#2{\left\langle#1\,#2\right\rangle}
\def\spb#1.#2{\left[#1\,#2\right]}
\def\lor#1.#2{\left(#1\,#2\right)}

\catcode`@=11  % Make @ letter.

\def\lsl{\not{\hbox{\kern-2.3pt $\ell$}}}
\def\ksl{\not{\hbox{\kern-2.3pt $k$}}}

\def\spa#1.#2{\left\langle#1\,#2\right\rangle}
\def\spb#1.#2{\left[#1\,#2\right]}
\def\lor#1.#2{\left(#1\,#2\right)}

\def\sand#1.#2.#3{%
  \left\langle\smash{#1}{\vphantom1}\right|{#2}%
  \left|\smash{#3}{\vphantom1}\right\rangle}
\def\sandp#1.#2.#3{%
  \left\langle\smash{#1}{\vphantom1}^{-}\right|{#2}%
  \left|\smash{#3}{\vphantom1}^{+}\right\rangle}
\def\sandpp#1.#2.#3{%
  \left\langle\smash{#1}{\vphantom1}^{+}\right|{#2}%
  \left|\smash{#3}{\vphantom1}^{+}\right\rangle}
\def\sandmm#1.#2.#3{%
  \left\langle\smash{#1}{\vphantom1}^{-}\right|{#2}%
  \left|\smash{#3}{\vphantom1}^{-}\right\rangle}
\def\sandpm#1.#2.#3{%
  \left\langle\smash{#1}{\vphantom1}^{+}\right|{#2}%
  \left|\smash{#3}{\vphantom1}^{-}\right\rangle}
\def\sandmp#1.#2.#3{%
  \left\langle\smash{#1}{\vphantom1}^{-}\right|{#2}%
  \left|\smash{#3}{\vphantom1}^{+}\right\rangle}

\def\nn{\nonumber}

\def\ap{\alpha'}
\def\ap{\alpha'}
\def\s(#1,#2){s_{#1,#2}}
\def\F#1#2(#3;#4;#5){ {}_{#1}{\rm F}_{#2}(#3;#4;#5)}
\def\frac#1#2{{#1\over#2}}

\def\text#1{{ #1}}

\def\oA#1{{\cal O}({\alpha'}^{#1})}

\newcommand\fverb{\setbox\fverbbox=\hbox\bgroup\verb}
\newcommand\fverbdo{\egroup\medskip\noindent%
			\fbox{\unhbox\fverbbox}\ }
\newcommand\fverbit{\egroup\item[\fbox{\unhbox\fverbbox}]}
\newbox\fverbbox

\title{Monodromy and Jacobi-like Relations for Color-Ordered
Amplitudes}

\author{N. E. J. Bjerrum-Bohr, Poul H. Damgaard and Thomas S{\o}ndergaard\\
Niels Bohr International Academy and DISCOVERY Center,\\ The Niels Bohr
Institute,
Blegdamsvej 17,\\ DK-2100 Copenhagen \O, Denmark,\\ {\tt email:} \{bjbohr;phdamg;tsonderg\}@nbi.dk}
\author{Pierre Vanhove\\
Institut des Hautes Etudes Scientifiques, Le Bois-Marie,\\
F-91440 Bures-sur-Yvette, France\\  and\\
CEA, DSM, Institut de Physique Th\'eorique, IPhT, CNRS, MPPU,\\
URA2306, Saclay, F-91191 Gif-sur-Yvette, France,\\ {\tt email:} pierre.vanhove@cea.fr}

\received{\today} 		%%
\accepted{\today}		%% These are for published papers.

\preprint{IPHT-10-030, IHES/P/10/08}
\abstract{
We discuss monodromy relations between different color-ordered
amplitudes in gauge theories. We show that Jacobi-like relations of Bern,
Carrasco and Johansson can be introduced in a manner that is compatible
with these monodromy relations. The Jacobi-like relations are not
the most general set of equations that satisfy this criterion.
Applications to supergravity amplitudes follow straightforwardly
through the KLT-relations. We explicitly show how
the tree-level relations give rise to non-trivial identities
at loop level.
}


\begin{document}

\section{Introduction}
One of the most striking aspects of string theory is the manner
in which it reorganizes the perturbative calculation of
amplitudes in the field theory limit. Perhaps the most
remarkable example of this is found in the Kawai-Lewellen-Tye
(KLT) relations~\cite{KLT} that link gauge field tree-level
amplitudes based on a non-Abelian gauge group to tree-level
amplitudes in perturbative gravity. As it is based on a
relationship between closed and open
strings~\cite{Green:1987sp}, it immediately yields an even
larger class of relations when considered in the context of
superstring theory: a whole set of relations between
supergravity and supersymmetry multiplets at tree level. For a
comprehensive discussion, see, {\it e.g.}, the review by
Bern~\cite{Bern:2002kj}. These relations are puzzling from the
point of view of field theory itself, although there are
attempts to see their origin at the Lagrangian
level~\cite{Bern:1999ji}.

Recently, three of the present authors have provided another
example of how string theory can be used to derive non-trivial
amplitude relations that hold even in the field theory limit,
although their origin remains mysterious
there~\cite{BjerrumBohr:2009rd}. The relations were conjectured
earlier by Bern-Carrasco-Johansson~\cite{Bern:2008qj}, and we
shall call them BCJ-relations in what follows. The peculiar
aspect in this case is that these BCJ-relations seemed to
follow from a new principle of Jacobi-like relations among
tree-level amplitudes~\cite{Bern:2008qj}, relations that hold
on-shell for four-point amplitudes~\cite{Zhu:1980sz}, but which
do {\em not} hold off-shell. Nevertheless, imposing these
Jacobi-like relations even above four-point amplitudes yields
correct amplitude relations. It was subsequently shown that
analogous amplitude relations can be derived for external
particles of the full ${\cal N}=4$
hypermultiplet~\cite{Sondergaard:2009za}, a result that indeed
also follows directly from the proof using superstring
theory~\cite{BjerrumBohr:2009rd}.

To understand the significance of a new set of amplitude
relations one needs to consider the factorial growth in $n$ for
color-ordered $n$-point amplitudes. For a tree-level $n$-point
amplitude $\cA_n$ with legs in the adjoint representation of,
say, $SU(N)$ gauge group, one defines the color-ordered
$n$-point amplitude $A_n(1,\ldots,n)$ through
\BE
\cA_n = g_{\rm YM}^{n-2}\hspace{-0.3cm}\sum_{\sigma\in S_n/\ZZ_n}
\hspace{-0.3cm} \mathrm{Tr}(T^{a_{\sigma(1)}}\cdots T^{a_{\sigma(n)}})\, A_n(\sigma(1,\ldots,n))\,,
\EE
where $g_{YM}$ is the coupling constant, and the $T$'s are group
generators of $SU(N)$. The relations we shall discuss all concern
the color-ordered amplitudes $A_n(1,\ldots,n)$. Of course, to
obtain cross sections, these must be ``dressed'' with the
appropriate color factors and summed. The shorter the sum,
the faster will routines work that do this sum automatically.
It is therefore not only of theoretical interest, but also
of great practical value to have exact relations
available among the color-order amplitudes. Because of cyclicity
of the amplitudes, the basis is not of size $n!$ but of size
$(n-1)!$ Additional non-trivial generic relations
known before the BCJ-relations were the following. Reflections:
\BE
A_n(1,\ldots,n) = (-1)^nA_n(n,n-1,\ldots,2,1)\,,
\EE
the photon decoupling relation
\BE
0 ~=~ \sum_{\sigma} A_n{\left(1,\sigma(2,\ldots,n)\right)}\,,
\EE
and the Kleiss-Kuijf relations~\cite{Kleiss:1988ne}
\begin{equation}\begin{split}
\hspace{0cm}A_n(\beta_1,\ldots,\beta_r,1,
\alpha_1,\ldots,\alpha_s,n)&=
(-1)^r\!\!\!\!\!\!\!\!\sum_{\sigma\subset{\rm OP}
\{\alpha\}\cup\{\beta^T\}}\!\!\!\!\!\!\!
A_n(1,\sigma,n)\,,
\end{split}\end{equation}
where the sum runs over the {\em ordered set of permutations}
that preserves the order within each set. Transposition on the
set $\{\beta\}$ means that order is reversed.

It was shown in ref.~\cite{Kleiss:1988ne,DelDuca:1999rs} that
these relations reduce the basis of amplitudes from $(n-1)!$ to
$(n-2)!$ The BCJ-relations reduce the basis down to $(n-3)!$ As
follows from the proof based on
monodromy~\cite{BjerrumBohr:2009rd}, no further reduction for
arbitrary $n$ will be possible. After imposition of the
BCJ-relations one has thus reached the minimal basis of
amplitudes.

In this paper we confront some of the questions that are raised
by the apparently valid imposition of Jacobi-like relations
among tree-level amplitudes. Given that the BCJ-relations have
now been {\em proven} based on
monodromy~\cite{BjerrumBohr:2009rd} a natural question is
whether the Jacobi-like relations, conversely, follow from the
BCJ-relations. Not unexpectedly, we find that this is not the
case. In fact, we find that a huge extension of these
Jacobi-like relations is possible~\footnote{ \footnotesize{In
the process of completing this manuscript a paper by H. Tye and
Y. Zhang~\cite{Tye:2010dd} appeared. They consider amplitude
relations from the viewpoint of heterotic string models. Some
of their results overlap with ours, in particular regarding the
existence of {\it extended} (or {\it generalized}) Jacobi
identities, which we discuss in sections 3 and 4.}}, still
leaving invariant the BCJ-relations.

The paper is organised as follows. In section 2, we briefly
review monodromy relations in string theory, and show how they
give rise to string theory generalizations of both the
Kleiss-Kuijf and BCJ-relations. Section 3 contains a discussion
of the connection between monodromy and Jacobi-like relations.
There are clearly some issues related to gauge symmetry, and we
choose in section 4 to consider this from the point of view of
string theory, which automatically imposes a specific gauge
choice. In section 5, we turn to gravity, and consider the
extended Jacobi-like identities in the light of KLT-relations.
All of these issues concern tree-level amplitudes only. In
section 6, we explore what these by now established tree-level
identities imply for loop amplitudes. A straightforward way to
attack this is through the use of cuts. We illustrate this in
the most simple case of one-loop amplitudes in $\mathcal{N}=4$
super Yang-Mills theory and comment on applications to theories
with less, or no, supersymmetry. Finally, section 7 contains
our conclusions. Some details about hypergeometric functions
are relegated to an appendix.

\section{Monodromy relations}\label{mono}
In this section we will briefly recall how to derive
monodromy relations for amplitudes through string theory. The color-ordered amplitudes on
the disc are given by~\cite{Green:1987sp}
\begin{align}
\label{eq:orderedBos}
\cA_n{(a_1,\ldots,a_n)}=\int\!\!\prod_{i=1}^{n}
dz_i\,{|z_{ab}\,z_{ac}\,z_{bc}|\over dz_adz_bdz_c}  \prod_{i=1}^{n-1}
H(x_{a_{i+1}}\!-\!x_{a_i})\!\!\!\!
\prod_{1\leq i<j\leq n}\!\!\!\!               \,
|x_i-x_j|^{2\alpha'k_i\cdot k_j}\, F_n\,,
\end{align}
with
\begin{equation}\begin{split}
\begin{array}{rllll}
 dz_i &=dx_i  & {\ \ \ \ \rm and \ \ \ \ }  z_{ij} & =x_i-x_j\ \ \  \ & {\rm \ \ for\ the\ bosonic\ case\ and\ } \\
dz_i &=dx_i d\theta_i & {\rm \ \ \ \ and \ \ \ \ }  z_{ij}& =x_i-x_j+\theta_i\theta_j\ \ \ \  & {\rm \ \ for \
the \ supersymmetric \ case\,.}
\end{array}
\end{split}\end{equation}
The ordering of the external legs is
enforced by the product of Heaviside functions such that
\begin{equation}
H(x)=\left\{\begin{array}{cc}
0 &\quad\quad x<0\,,\\
1 &\quad\quad x\geq0\,.\end{array}
\right.\end{equation}
The M\"obius $SL(2,\mathbb{R})$
invariance requires one to fix the position
of three points denoted $z_a$, $z_b$ and $z_c$. A traditional
choice is $x_1=0$, $x_{n-1}=1$ and $x_n=+\infty$, supplemented
by the condition $\theta_{n-1}=\theta_n=0$ in the superstring
case.

The helicity dependence of the external states is contained in
the $F_n$ factor. For tachyons $F_n=1$. For $n$
gauge bosons with polarization vectors $h_i$ one has
\begin{equation}
F_n=\exp\left(-\sum_{i\neq j}\bigg( {\sqrt{\alpha'} (h_i\cdot
k_j)\over (x_i-x_j)}-2{(h_i\cdot
  h_j)\over   (x_i-x_j)^2}\bigg)\right)\bigg|_{\rm  multilinear~in~{\it h}_i}\,,
\end{equation}
for  the bosonic string. For the superstring $F_n$ reads (the $\eta_i$ are anticommuting
variables)
\begin{equation}
F_n=  \int   \prod_{i=1}^n
d\eta_i  \, \exp\left( - \sum_{i\neq j}\bigg({ \eta_i
\sqrt{\alpha'} (\theta_i-\theta_j)(h_i\cdot k_j)-
\eta_i\eta_j(h_i\cdot
  h_j)\over (x_i-x_j+\theta_i\theta_j)}\bigg)\right)\,.
\label{super}\end{equation}

We start with a review of the
monodromy relations that appear at four
points~\cite{Plahte:1970wy,Dotsenko:1984nm,BjerrumBohr:2009rd}.
For simplicity, we phrase the discussion in terms of tachyon
amplitudes. With the choice  $x_1=0$, $x_3=1$ and
$x_4=+\infty$, all three different color-ordered amplitudes
$\cA{(i,j,k,l)}$ are given by the same integrand
$$|x_2|^{2\alpha'\,k_1\cdot k_2}|1-x_2|^{2\alpha'\,k_2\cdot k_3}\,,$$
but with $x_2$ integrated over different domains:
\begin{eqnarray}
\label{e:A1234}\!\!\!\!\!\cA_4{(1,2,3,4)}&=&\!\int_0^1 \!dx \,\, x^{2\alpha'\,k_1\cdot   k_2}
(1-x)^{2\alpha'\,k_2\cdot k_3}\,,\\
\!\!\!\!\!\cA_4{(1,3,2,4)}\!&=&\!\int_1^\infty \!\!\!dx\,\, x^{2\alpha'\, k_1\cdot k_2}
(x-1)^{2\alpha'\,k_2\cdot k_3}\,,\\
\!\!\!\!\!\cA_4{(2,1,3,4)}\!&=&\!\int_{-\infty}^0\!\!\!\!\! dx\,(-x)^{2\alpha'\, k_1\cdot k_2}
(1-x)^{2\alpha'\,k_2\cdot k_3}\,.
\end{eqnarray}
We indicate the contour integration from 1 to $+\infty$ in fig.~\ref{fig1}.\\
\vspace{-0.6cm}
\FIGURE[h]{
\centering
\includegraphics[width=12cm]{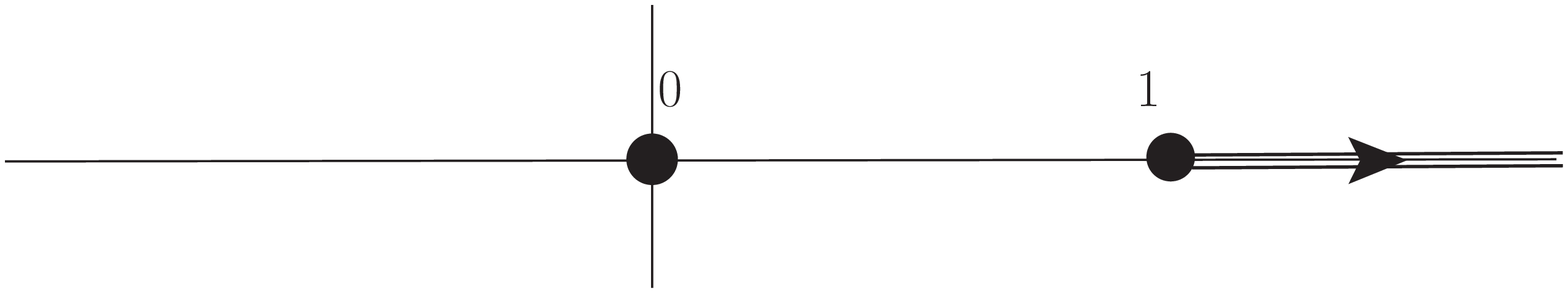}
\label{fig1} \caption{\sl The contour of integration from 1 to
$+\infty$. }}\vspace{-0.3cm}

\noindent Under the assumption that $\alpha'\,k_i\cdot  k_j$ is
complex and has a negative real part, we are allowed to deform
the region of integration so that instead of integrating
between from 1 to $+\infty$ on the real axis we integrate
either on a contour slightly above or below the real axis. By a
deformation of each of the contours, one can convert the
expression into an integration from $-\infty$  to 1. One needs
to include the appropriate phases each time $x$ passes through
$y=0$ or $y=1$ (when rotating the contours),
\begin{equation}
\label{eq:phaseP}
(x-y)^{\alpha}= (y-x)^{\alpha}\times
\begin{cases}
e^{+ i\pi \,\alpha}&\textrm{for~clockwise~rotation}\,, \cr
e^{- i\pi \,\alpha}&\textrm{for~counterclockwise~rotation}\,.
\end{cases}\nonumber
\end{equation}
One can thus deform the integration region in
two equivalent ways ${\cal I}^+$ and ${\cal I}^-$, see fig.~\ref{fig3}. \\
\vspace{-0.1cm}
\FIGURE[h]{
\centering
\includegraphics[width=11cm]{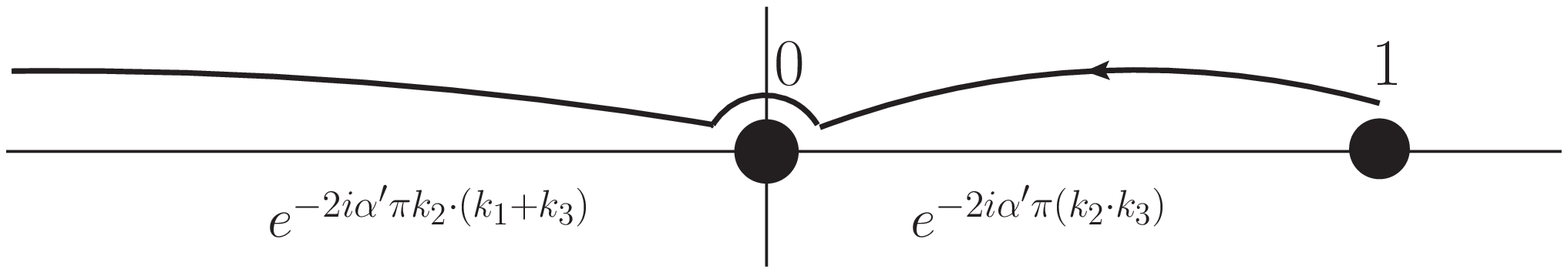}
\label{fig2} }
\FIGURE[h]{
\centering
\includegraphics[width=11cm]{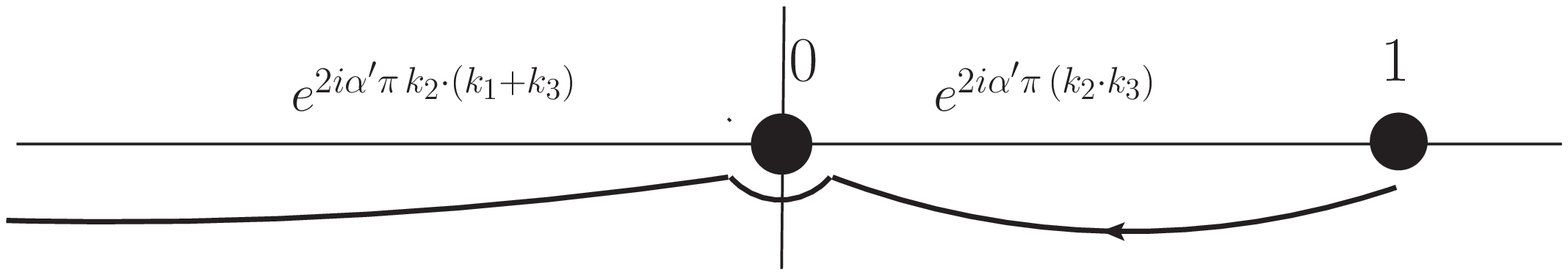}
\label{fig3}\caption{The contours ${\cal I}^+$ and ${\cal I}^-$. } }

\FIGURE[h!!!!!!!]{ \centering
\includegraphics[width=11cm]{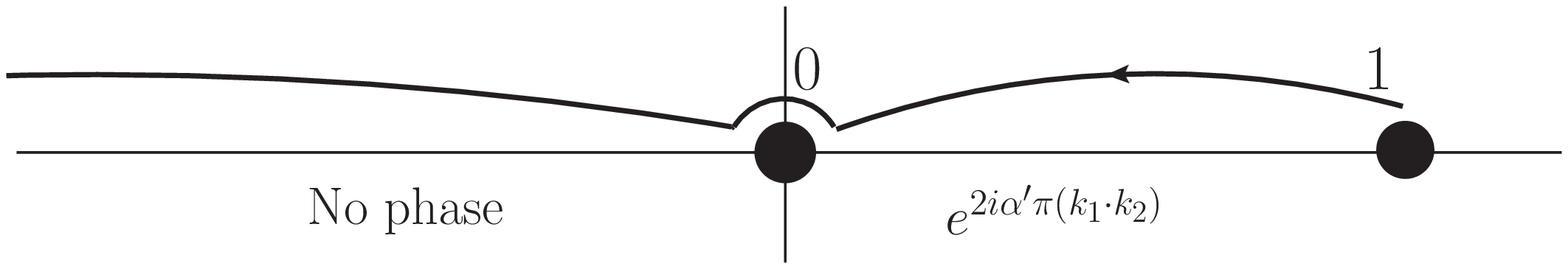}}
\FIGURE[h!!!!!!!!!!]{ \centering
\includegraphics[width=11cm]{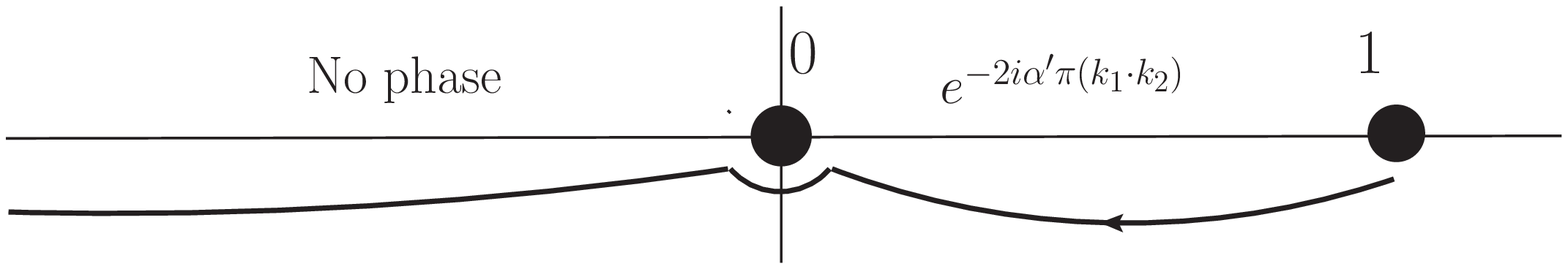}
\label{fig5}\caption{The contours ${\cal I}^+$ and ${\cal I}^-$
after multiplying with phases $e^{2i\alpha'\pi k_2\cdot
(k_1+k_3)}$ and $e^{-2i\alpha'\pi k_2\cdot(k_1+ k_3)}$. } }

We have ${\cal I}^+ = {\cal I}^- = \cA_4{(1,3,2,4)}$. If now
${\cal I}^+$ is multiplied by $e^{2i\alpha'\pi k_2\cdot
(k_1+k_3)}$ and ${\cal I}^-$ by $e^{-2i\alpha'\pi k_2\cdot(k_1+
k_3)}$ we get for the contours as illustrated in
fig.~\ref{fig5}. We thus have ${\cal I}^+\,e^{2i\alpha'\pi
k_2\cdot (k_1+k_3)} -{\cal I}^-\,e^{-2i\alpha'\pi k_2\cdot
(k_1+k_3)}=2i\,\cA_4{(1,3,2,4)}\,\sin(2\alpha'\pi\,k_2\cdot
(k_1+k_3))$. However, the contour obtained after subtracting
these two contours can also be interpreted as in
fig.~\ref{fig7}. This is equal to
$-2i\,\cA_4{(1,2,3,4)}\,\sin(2\alpha'\pi k_1\cdot k_2)$. In
this way we arrive at the following monodromy relation:
$\sin(2\pi \alpha' k_1\cdot k_2)
\cA_4{(1,2,3,4)}=\sin(2\pi\alpha' k_2\cdot k_4)
\cA_4{(1,3,2,4)}$
where we have used momentum conservation and the on-shell
condition. For other external states of higher spin, the
integrals change appropriately to restore the identities
(including sign factors for the fermionic statistics of
half-integer spins).

\FIGURE[h]{ \centering
\includegraphics[width=11cm]{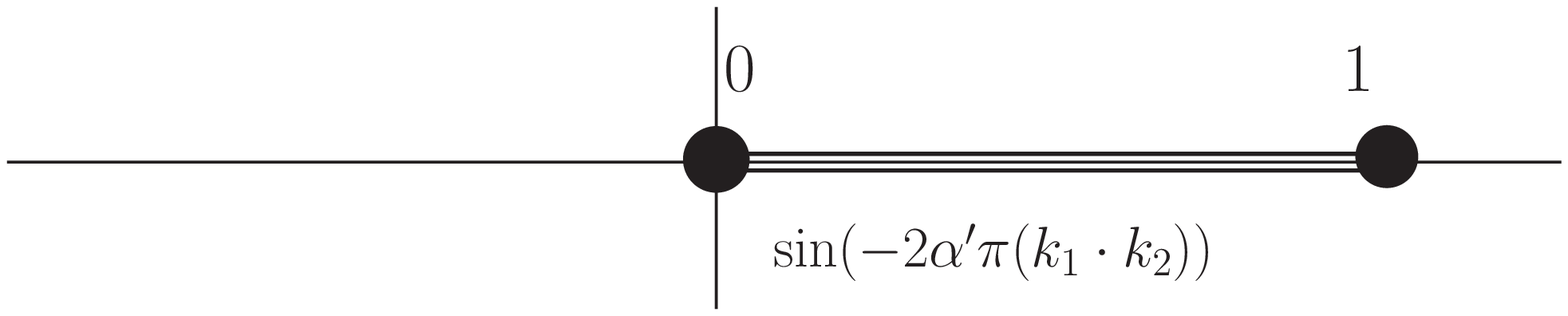}
\label{fig7}\caption{Another interpretation of the two contours.} }
\noindent

By deforming the contour of integration of $\cA_4(2,1,3,4)$ one
finds in an equivalent fashion: $\sin(2\pi \alpha'  k_2\cdot
k_3)\cA_4{(1,2,3,4)}= \sin(2\pi\alpha' k_2\cdot k_4)
\cA_4{(2,1,3,4)}$. This implies that all the amplitudes can be
related to the $\cA_4{(1,2,3,4)}$
\begin{equation}
\label{e:relationFour}
\begin{split}
\cA_4{(1,3,2,4)}&={\sin(2\pi\alpha'k_1\cdot k_2)\over
\sin(2\pi\alpha'k_2\cdot k_4)} \,\cA_4{(1,2,3,4)}\,,\cr
\cA_4{(1,3,4,2)}=\cA_4{(2,1,3,4)}
&={\sin(2\pi\alpha'k_2\cdot k_3)\over \sin(2\pi\alpha'k_2\cdot k_4)}
\,\cA_4{(1,2,3,4)}\,.
\end{split}\end{equation}\vskip-6pt\noindent
Taking the limit $\alpha'\rightarrow  0$, we get  the following
relations between the field theory amplitudes
\begin{equation}
\label{e:relationFour}
\begin{split}
A_4{(1,3,2,4)}&={k_1\cdot k_2\over k_2\cdot k_4} \,A_4{(1,2,3,4)}\,,
\cr
A_4{(1,3,4,2)}=A_4{(2,1,3,4)}&
={k_2\cdot k_3\over k_2\cdot k_4} \,A_4{(1,2,3,4)}\,.\
\end{split}\hspace{1.2cm}\end{equation}
The string theory relations can immediately be checked to hold
based on the explicit string amplitude expression. In the low
energy limit, the corresponding relations
(\ref{e:relationFour}) coincide with those of
ref.~\cite{Bern:2008qj}.

As shown in ref.~\cite{BjerrumBohr:2009rd}, one has the
following $n$-point amplitude relations:
\begin{align}\label{stringKK}
&{\cal A}_n(\beta_1,\ldots,\beta_r,1,\alpha_1,\ldots,\alpha_s,n)
=(-1)^r \nonumber \\
&\phantom{AAAAAAAAA}\times \Re{\rm e}\Big[\!\!\!\!\!\!
\prod_{1\leq i <j \leq r}\!\!\!\!e^{2i\pi\alpha'(k_{\beta_i}\cdot
k_{\beta_j})}\!\!\!\!\!\!\!\!\!
\sum_{\sigma\subset{\rm OP}
\{\alpha\}\cup\{\beta^T\}}\!\prod_{i=0}^s
\prod_{j=1}^r\! e^{(\alpha_i,\beta_j)}{\cal A}_n(1,\sigma,n)\Big]\,,
\end{align}
\begin{equation}\begin{split}\label{stringBern}
0=\hspace{-0cm}\Im{\rm m}\Big[\!\!\!\!
\prod_{1\leq i <j \leq r}\!\!\!
\!e^{2i\pi\alpha'(k_{\beta_i}\cdot k_{\beta_j})}\!\!\!\!\!\!\!\!\!
\sum_{\sigma\subset{\rm OP}
\{\alpha\}\cup\{\beta^T\}}\!\prod_{i=0}^s\prod_{j=1}^r
\!e^{(\alpha_i,\beta_j)}\cA_n(1,\sigma,n)\Big]\,,\
\end{split}\end{equation}\noindent
with\\
\begin{center}
$e^{({\alpha,\beta})}$ \ $\equiv$ \  $\bigg\{$\begin{tabular}{ccccc}
 & & $e^{2i\pi\alpha'(k_\alpha\cdot
k_\beta)}$ & if & $x_{\beta} >  x_{\alpha}$\,,\\
&& $1$ &  & otherwise.
\end{tabular}
\end{center}
In these equations $\alpha_0$ denotes the leg 1 at point 0.

These string theory amplitude relations reduce in the field
theory limit $\alpha' \to 0$ to the
Kleiss-Kuijf~\cite{Kleiss:1988ne,DelDuca:1999rs} and
BCJ-relations~\cite{Bern:2008qj}, respectively.

Explicitly, using~\eqref{stringBern} as well as momentum conservation,
the five-point amplitude gives rise to the following four
independent relations
\begin{align}
&0=\cSS{k_3}{k_1+k_2}\cA_5(1,2,3,4,5)-\cSS{k_3}{k_5}\cA_5(1,2,4,3,5)+\cSS{k_1}{k_3}\cA_5(1,3,2,4,5)\,,\nonumber \\
&0=\cSS{k_3}{k_2+k_5}\cA_5(1,4,3,2,5)-\cSS{k_1}{k_3}\cA_5(1,3,4,2,5)+\cSS{k_3}{k_5}\cA_5(1,4,2,3,5)\,,\nonumber \\
&0=\cSS{k_4}{k_2+k_5}\cA_5(1,3,4,2,5)-\cSS{k_1}{k_4}\cA_5(1,4,3,2,5)+\cSS{k_4}{k_5}\cA_5(1,3,2,4,5)\,,\nonumber \\
&0=\cSS{k_2}{k_4+k_5}\cA_5(1,3,2,4,5)-\cSS{k_1}{k_2}\cA_5(1,2,3,4,5)+\cSS{k_2}{k_5}\cA_5(1,3,4,2,5)\,. \label{stringBernFive}
\end{align}
Here we have used the notation $\cSS{p}{q}\equiv
\sin(2\alpha'\pi\,p\cdot q)$.
There are of course various ways of writing these monodromy
relations, but they reduce to just four independent equations.
One can immediately verify these relations from the explicit form
of the tree amplitudes in string theory given
by~\cite{Medina:2002nk,Stieberger:2006te,Stieberger:2009hq,Cheung:2010vn}.
In the
field theory limit they reduce to relations that are equivalent to those
discussed in ref.~\cite{Bern:2008qj}.

\section{Jacobi-like identities}
The field theory limit of the monodromy relations were originally
conjectured on the basis of an observation
for the four-point gluon amplitudes~\cite{Bern:2008qj}.
We start by briefly reviewing the argument.
\subsection{The four-point case}
At four points, the photon decoupling identity reads
\begin{eqnarray}
A_4(1,2,3,4) + A_4(2,1,3,4) + A_4(2,3,1,4) = 0\,.
\end{eqnarray}
It holds independently of polarization and external on-shell
momenta. The natural way this identity can be satisfied is
through
\begin{eqnarray}
A_4(1,2,3,4) + A_4(2,1,3,4) + A_4(2,3,1,4) = \chi (s+t+u) = 0\,,
\label{phodecoup}
\end{eqnarray}
with $\chi$ being a common factor~\footnote{\footnotesize{We
will discuss the explicit expression for $\chi$ in the case of
vector particles in section 4.}}.

In the amplitude $A_4(1,2,3,4)$ both pairs of legs (1,2) and
(1,4) are adjacent, and we should thus treat the $s$ and $t$
factors on the same footing. The contribution of this color
ordering to eq.~\eqref{phodecoup} must therefore be
\begin{eqnarray}
A_4(1,2,3,4) = -\chi(s+t) = \chi u\,.
\end{eqnarray}
Likewise, one is led to
\begin{eqnarray}
A_4(2,1,3,4) = \chi t, \qquad A_4(2,3,1,4) = \chi s\,.
\end{eqnarray}
Eliminating $\chi$ one obtains
\begin{eqnarray}
&tA_4(1,2,3,4) = uA_4(2,1,3,4),
\qquad sA_4(1,2,3,4) = uA_4(2,3,1,4),& \nonumber \\
&sA_4(2,1,3,4) = tA_4(2,3,1,4)\,.&
\label{threerel}
\end{eqnarray}
These are of course just the monodromy relations
eq.~\eqref{e:relationFour}. To proceed further, one can
parameterize the three subamplitudes in terms of their possible
pole structures and unspecified numerators
\begin{align}
A_4(1,2,3,4) &= \frac{n_s}{s} + \frac{n_t}{t}\,, \label{first} \\
A_4(2,1,3,4) &= -\frac{n_u}{u} - \frac{n_s}{s}\,, \label{second} \\
A_4(2,3,1,4) &= -\frac{n_t}{t}+\frac{n_u}{u}\,. \label{third}
\end{align}
It follows from (\ref{threerel}) that $n_u-n_s+n_t = 0$. This
resembles the Jacobi identity for the associated color factors.
Bern, Carrasco and Johansson~\cite{Bern:2008qj} took as
hypothesis that this can be extended iteratively for general
$n$-point amplitudes. This is equivalent to assuming that one
can choose a parametrization in which Jacobi relations for
numerator factors can be imposed in one-to-one correspondence
with the genuine Jacobi identities for the color factors.
Imposing this hypothesis gets quite involved as $n$ grows, but
it can be carried through systematically; for details see
ref.~\cite{Bern:2008qj}. This leads to the
BCJ-relations~\cite{Bern:2008qj}. The same principle can be
used to generate relations for scalar and fermionic matter in
the adjoint representation~\cite{Sondergaard:2009za}. We of
course now understand that this is because the monodromy
relations hold for the full ${\cal N} = 4$ supermultiplet in
four dimensions~\cite{BjerrumBohr:2009rd}.

Since the BCJ-relations have been proven~\cite{BjerrumBohr:2009rd}, one
would like to understand the meaning of these Jacobi-like identities
for the numerators. In the four-point case the identities are
exact, but only on-shell~\cite{Zhu:1980sz}.
Even if the theory in question had only three-point
vertices (which it does not) so that all $n$-point
tree-level amplitudes for $n \geq 5$ could be constructed by gluing
three-point vertices on a four-point function (thus having at
least one leg off-shell on all four-point sub-diagrams),
this would represent a puzzle. How can this starting
point then lead to correct amplitude identities?

\subsection{Generalized Jacobi-like relations}
To see what is going on it suffices to focus on the 5-point case. We
will simply derive exactly what follows directly from the
field theory BCJ-relations
when expressed in terms of the pertinent set of poles for each
color-ordered amplitude. We use the parametrization
\begin{equation}\begin{split}\label{partialfive1}
A_5(1,2,3,4,5) & = \frac{n_1}{s_{12}s_{45}} +
\frac{n_2}{s_{23}s_{51}} + \frac{n_3}{s_{34}s_{12}} +
\frac{n_4}{s_{45}s_{23}} + \frac{n_5}{s_{51}s_{34}}\,,
\end{split}\end{equation}
\begin{equation}\begin{split}\label{partialfive2}
A_5(1,4,3,2,5) &= \frac{n_6}{s_{14}s_{25}} +
\frac{n_5}{s_{43}s_{51}} + \frac{n_7}{s_{32}s_{14}} +
\frac{n_8}{s_{25}s_{43}} + \frac{n_2}{s_{51}s_{32}}\,,
\end{split}\end{equation}
\begin{equation}\begin{split}\label{partialfive3}
A_5(1,3,4,2,5) &= \frac{n_9}{s_{13}s_{25}} -
\frac{n_5}{s_{34}s_{51}} + \frac{n_{10}}{s_{42}s_{13}} -
\frac{n_8}{s_{25}s_{34}} + \frac{n_{11}}{s_{51}s_{42}}\,,
\end{split}\end{equation}
\begin{equation}
\begin{split}\label{partialfive4}
A_5(1,2,4,3,5) &= \frac{n_{12}}{s_{12}s_{35}} +
\frac{n_{11}}{s_{24}s_{51}} - \frac{n_3}{s_{43}s_{12}} +
\frac{n_{13}}{s_{35}s_{24}} - \frac{n_5}{s_{51}s_{43}}\,,
\end{split}\end{equation}
\begin{equation}\begin{split}\label{partialfive5}
A_5(1,4,2,3,5) &= \frac{n_{14}}{s_{14}s_{35}} -
\frac{n_{11}}{s_{42}s_{51}} - \frac{n_7}{s_{23}s_{14}} -
\frac{n_{13}}{s_{35}s_{42}} - \frac{n_2}{s_{51}s_{23}}\,,
\end{split}\end{equation}
\begin{equation}\begin{split}\label{partialfive6}
A_5(1,3,2,4,5) &= \frac{n_{15}}{s_{13}s_{45}} -
\frac{n_2}{s_{32}s_{51}} - \frac{n_{10}}{s_{24}s_{13}} -
\frac{n_4}{s_{45}s_{32}} - \frac{n_{11}}{s_{51}s_{24}}\,.
\end{split}\end{equation}
This can be easily illustrated by diagrams involving only
anti-symmetric three-vertices. However, since the coefficients
$n_i$ may depend on the kinematic variables (and thus cancel
poles) there is no assumption of only three-vertices here. The
listed subamplitudes are related through the monodromy
relations in the field limit of~\eqref{stringBernFive}, {\it
i.e.},
\begin{align}
&0=(s_{13}+s_{23})A_5(1,2,3,4,5)-s_{35}A_5(1,2,4,3,5)+
s_{13}A_5(1,3,2,4,5)\,, \label{one} \\
&0=(s_{23}+s_{35})A_5(1,4,3,2,5)-
s_{13}A_5(1,3,4,2,5)+s_{35}A_5(1,4,2,3,5)\,, \label{two} \\
&0=(s_{24}+s_{45})A_5(1,3,4,2,5)-s_{14}A_5(1,4,3,2,5)+
s_{45}A_5(1,3,2,4,5)\,, \label{three} \\
&0=(s_{24}+s_{25})A_5(1,3,2,4,5)-s_{12}A_5(1,2,3,4,5)+
s_{25}A_5(1,3,4,2,5)\,. \label{four}
\end{align}

Plugging the expressions for the amplitudes in terms of the
$n_i$'s into~\eqref{one}--\eqref{four} we immediately obtain:
\begin{enumerate}
 \item From~\eqref{one}
\begin{equation}
0 =\frac{n_4-n_1+n_{15}}
{s_{45}} -\frac{n_{10}-n_{11}+n_{13}}{s_{24}}-
\frac{n_3-n_1+n_{12}}{s_{12}}-\frac{n_5-n_2+n_{11}}{s_{51}}\,,
\end{equation}
\item From~\eqref{two}
\begin{align}
0 = \frac{n_7-n_6+n_{14}}{s_{14}}-
\frac{n_{10}-n_{11}+n_{13}}{s_{24}}-\frac{n_8-n_6+n_9}{s_{25}}
-\frac{n_5-n_2+n_{11}}{s_{51}}\,,
\end{align}
\item From~\eqref{three}
\begin{align}
0 = \frac{n_{10}-n_9+n_{15}}{s_{13}} + \frac{n_5-n_2+n_{11}}{s_{51}}
-\frac{n_4-n_2+n_7}{s_{23}}+\frac{n_8-n_6+n_9}{s_{25}}\,,
\end{align}
\item From~\eqref{four}
\begin{align}
0 = \frac{n_{4}-n_1+n_{15}}{s_{45}} - \frac{n_{10}-n_9+n_{15}}{s_{13}}
-\frac{n_5-n_2+n_{11}}{s_{51}}-\frac{n_3-n_5+n_8}{s_{34}}\,.
\end{align}
\end{enumerate}
We thus see that the BCJ-relations can be written as kind of
extended Jacobi identities when expressed
in terms of the numerators. Let us simplify the notation a bit by
denoting the nine numerator combinations as
\begin{eqnarray}
\begin{array}{lclcl}
X_1 \equiv n_3-n_5+n_8\,, & & X_2 \equiv n_3-n_1+n_{12}\,, & & X_3 \equiv n_4-n_1+n_{15}\,, \\
X_4 \equiv n_4-n_2+n_7\,, & & X_5 \equiv n_5-n_2+n_{11}\,, & & X_6 \equiv n_7-n_6+n_{14}\,, \\
X_7 \equiv n_8-n_6+n_9\,, & & X_8 \equiv n_{10}-n_9+n_{15}\,, & & X_9 \equiv n_{10}-n_{11}+n_{13}\,.
\end{array}     \label{Xsol}
\end{eqnarray}
Our four equations then take the form
\begin{align}
&0 =\frac{X_3}{s_{45}} -\frac{X_9}{s_{24}}-
\frac{X_2}{s_{12}}-\frac{X_5}{s_{51}}\,,
\label{mone} \\
&0 =\frac{X_6}{s_{14}}-\frac{X_9}{s_{24}}-
\frac{X_7}{s_{25}}-\frac{X_5}{s_{51}}\,,
\label{mtwo} \\
&0 = \frac{X_8}{s_{13}} + \frac{X_5}{s_{51}}
-\frac{X_4}{s_{23}}+\frac{X_7}{s_{25}}\,,
\label{mthree} \\
&0 = \frac{X_3}{s_{45}} - \frac{X_8}{s_{13}}-
\frac{X_5}{s_{51}}-\frac{X_1}{s_{34}}\,.
\label{mfour}
\end{align}
These four equations describe the general constraints on the
numerator factors dictated by the monodromy relations at five
points. As long as these equations are satisfied we have
numerator identities leading to eq.~\eqref{one}--\eqref{four}.
Of course, the simplest solution is to put all $X_i = 0$, but
this is clearly not the most general solution.
\subsection{Reparametrization invariance}
To make the amount of freedom one has in the above parametrization of
subamplitudes more clear,
let us write the most general solution by means of
five arbitrary functions $f_1$, $f_2$, $f_3$, $f_4$ and $f_5$
\begin{align}
X_1 \equiv s_{34}f_1, \qquad X_2 \equiv s_{12}f_2\,,
\qquad X_3 \equiv s_{45}f_3, \qquad X_4 \equiv s_{23}f_4\,,
 \qquad X_5 \equiv s_{15}f_5\,,
\end{align}
{\it i.e.} from eq.~\eqref{mone}--\eqref{mfour}
\begin{eqnarray}
\begin{array}{lclcl}
X_1 \equiv s_{34}f_1\,, & & X_2 \equiv s_{12}f_2\,, & & X_3 \equiv s_{45}f_3\,, \\
X_4 \equiv s_{23}f_4\,, & & X_5 \equiv s_{15}f_5\,, & & X_6 = s_{14}(f_1-f_2+f_4)\,, \\
X_7 = s_{25}(f_1-f_3+f_4)\,, & &
X_8 = s_{13}(f_3-f_1-f_5)\,, & & X_9 = s_{24}(f_3-f_2-f_5)\,.
\end{array}
\label{generalsol}
\end{eqnarray}
Note that we have used the canonical set of kinematic variables (generalized
Mandelstam variables for the 5-point case)
$s_{12},s_{23},s_{34},s_{45},s_{51}$ in our definition of the $f_i$.
The $s_{ij}$ occuring in the expression for $X_6$, $X_7$, $X_8$ and $X_9$
are related to this canonical set by
\begin{eqnarray}
s_{14} = s_{23}-s_{15}-s_{45}, \qquad s_{25} = s_{34}-s_{12}-s_{15}\,, \nonumber \\
s_{13} = s_{45}-s_{12}-s_{23}, \qquad s_{24} = s_{15}-s_{23}-s_{34}\,.
\end{eqnarray}

The freedom we have to generalize the solution, {\it i.e.}
eq.~\eqref{generalsol}, is not just related to gauge degrees or
the freedom to absorb contact terms. It can be seen as the
trivial freedom to add a ``zero'' to the subamplitude and
forcing it into a parametrization of the form
eq.~\eqref{partialfive1}--\eqref{partialfive6}.

As a simple example, imagine that we add $0 = g-g$ to
eq.~\eqref{partialfive1}, with $g$ being an arbitrary function.
We can then absorb the $g$'s in $n_1$ and $n_3$, {\it i.e.}
\begin{align}
A_5(1,2,3,4,5) &=
\frac{(n_1+s_{12}s_{45}g)}{s_{12}s_{45}} + \frac{n_2}{s_{23}s_{51}}
+ \frac{(n_3-s_{34}s_{12}g)}{s_{34}s_{12}} +
\frac{n_4}{s_{45}s_{23}} + \frac{n_5}{s_{51}s_{34}}\,.
\end{align}
In no other amplitude than $A_5(1,2,3,4,5)$ does $n_1$ appear,
however, $n_3$ appears in eq.~\eqref{partialfive4} so we add $0
= g-g$ to the amplitude, and absorb in the following way:
\begin{align}
A_5(1,2,4,3,5) &= \frac{(n_{12}-s_{12}s_{35}g)}{s_{12}s_{35}} +
\frac{n_{11}}{s_{24}s_{51}}
- \frac{(n_3-s_{34}s_{12}g)}{s_{43}s_{12}} +
\frac{n_{13}}{s_{35}s_{24}} - \frac{n_5}{s_{51}s_{43}}\,.
\end{align}
We have thereby redefined $n_1$, $n_3$ and $n_{12}$
\begin{align}
n_1 \quad &\rightarrow \quad n_1+s_{12}s_{45}g\,, \\
n_3 \quad &\rightarrow \quad n_3-s_{34}s_{12}g\,, \\
n_{12} \quad &\rightarrow \quad n_{12}-s_{12}s_{35}g\,,
\end{align}
which changes $X_1$, $X_2$ and $X_3$
\begin{align}
X_1 = s_{34}f_1 \quad &\rightarrow
\quad s_{34}(f_1-s_{12}g) \equiv s_{34}f_1'\,, \\
X_2 = s_{12}f_2\quad &\rightarrow
\quad s_{12}(f_2-(s_{45}+s_{34}+s_{35})g) =
s_{12}(f_2-s_{12}g) \equiv s_{12}f_2'\,, \\
X_3 = s_{45}f_3 \quad &\rightarrow \quad s_{45}(f_3-s_{12}g) \equiv s_{45}f_3'\,,
\end{align}
and we now have
\begin{eqnarray}
\begin{array}{lclcl}
X_1 = s_{34}f_1'\,, & & X_2 = s_{12}f_2'\,, & & X_3 = s_{45}f_3'\,, \\
X_4 = s_{23}f_4\,, & & X_5 = s_{15}f_5\,, & & X_6 = s_{14}(f_1'-f_2'+f_4)\,, \\
X_7 = s_{25}(f_1'-f_3'+f_4)\,, & & X_8 =
s_{13}(f_3'-f_1'-f_5)\,, & & X_9 = s_{24}(f_3'-f_2'-f_5)\,.
\end{array}
\end{eqnarray}
This trivial addition of zeros to the amplitudes illustrates
the fact that we can
find many different representations of the
numerators, all of which are perfectly consistent with the monodromy relations.
The freedom is that of general reparametrizations of the amplitude and
not just gauge symmetry.

%%%%%%%%%%%%%%%%%%%%%%%%%%%%%%%%%%%%%%%%%%%%%%%%%%%%%%%%%%%%%%%%%%%%%%%%%%%%%
\section{String amplitudes}
%%%%%%%%%%%%%%%%%%%%%%%%%%%%%%%%%%%%%%%%%%%%%%%%%%%%%%%%%%%%%%%%%%%%%%%%%%%%%
Let us consider tree-level open string amplitudes in
superstring theory. We have already given the needed formulas
in section~\ref{mono}. We first focus on the color-ordered
four-point amplitude for vector particles
\begin{equation}\label{eFp}
{\cal A}_4^\sigma =  \int_{D_\sigma} dz_2\, |z_2|^{2\alpha' \,k_1\cdot
k_2}   |1-z_2|^{2\alpha'\,k_2\cdot   k_3}\,  \tilde F_4(z_2)\,,
\end{equation}
where the domain of integration $D_\sigma$ for each color
ordering are given by $D_{1234} =\{0\leq z_2\leq 1\}\,,\
D_{1324} =\{1\leq z_2\}\,, \ D_{2134} =\{z_2\leq 0\}\,. $
Expanding the function $\tilde F_4$ in~(\ref{super})
leads\footnote{This  can be derived  with a  very tedious
expansion~\cite{Oprisa:2005wu} of the expression in
eq.~(\ref{super}). The simplicity of the expansion appears
naturally in the pure spinor
formalism~\cite{Berkovits:2000fe,Mafra:2009bz}. The tilde on
$F_n$ indicates that we have fixed the three conformal points
in the expression.} to
\begin{equation}
\label{e:fsusy}
\tilde F_4(y)= {a_1\over y}+{b_1\over y-1}\,,
\end{equation}
where $a_1$ and $b_1$ are  expressed in terms of the
polarizations and the momenta.  Their expressions are
particularly long but there is a relation between the two
coefficients
\begin{equation}\label{e:t8F}
s\, b_1- t\, a_1=\ap\, t_8^{m_1\cdots m_8}F^1_{m_1m_2}F^2_{m_3m_4} F^3_{m_5m_6} F^4_{m_7m_8}\,,
\end{equation}
where  $F^i$  are  the  field-strengths corresponding to the
external legs. The tensor $t_8$ is contracting  the Lorentz
indices as defined in appendix~9.A of~\cite{Green:1987sp} (it
is common to define $\chi= t_8^{m_1\cdots
m_8}F^1_{m_1m_2}F^2_{m_3m_4}F^3_{m_5m_6} F^4_{m_7m_8}/(stu)$).
The quantity $a_1$ and $b_1$ are  not gauge invariant but the
combination in~(\ref{e:t8F}) is gauge invariant.

For the four-point color-ordered amplitudes we find
\begin{eqnarray}\label{eAfourInt}
{\cal A}_4(1,2,3,4)&=&      \Phi_{2,1}( \ap\,s,\ap\,t)\,\left(-{a_1\over \ap\,s}+{b_1\over \ap\,t}\right)\,,\\
{\cal A}_4(1,3,2,4)&=&      \Phi_{2,1}( \ap\,u,\ap\,t)\,\left(-{a_1+b_1\over \ap\,u}-{b_1\over \ap\,t}\right)\,,\\
{\cal A}_4(2,1,3,4)&=&      \Phi_{2,1}( \ap\,s,\ap\,u)\,\left({a_1\over \ap\,s}+{a_1+b_1\over \ap\,u}\right)\,,
\end{eqnarray}
where we introduced the hypergeometric functions
\begin{equation}\label{e:Hyp21}
\Phi_{2,1}(\ap s,\ap t)\equiv\F21(-\ap\,s,\ap\,t;1-\ap\,s;1)= {\Gamma(1-\alpha's)\Gamma(1-\alpha't)\over
\Gamma(1+\alpha'u)}\,.
\end{equation}
In the convention of BCJ~\cite{Bern:2008qj},
\begin{equation}
  n_s= -a_1/\ap, \qquad n_t=-b_1/\ap, \qquad n_u=-(a_1+b_1)/\ap\,,
\end{equation}
we immediately obtain the exact relation $n_u=n_t-n_s$.

%%%%%%%%%%%%%%%%%%%%%%%%%%%%%%%%%%%%%%%%%%%%%%%%%%%%%%%%%%%%%%%%%%%%%%%%%%%%%
\subsection{Five points}
%%%%%%%%%%%%%%%%%%%%%%%%%%%%%%%%%%%%%%%%%%%%%%%%%%%%%%%%%%%%%%%%%%%%%%%%%%%%%

Let us now consider the five point amplitude. Having  fixed the
position vertex operators at positions $z_1=0$, $z_4=1$ and
$z_5=\infty$, the integrand takes the compact
form~\cite{Mafra:2009bz}

\begin{equation}\label{expre}
{\cal A}_5^\sigma = \int_{D_\sigma}  dz_2 dz_3 \prod_{i <
j}|z_{ij}|^{2\ap k_i\cdot k_j}
\Big[ {A\over z_{12}z_{13}} + {B\over z_{23}z_{24}}
+ {C\over z_{12}z_{34}} +{D\over z_{24}z_{34}}
+ {E\over z_{23}z_{13}}  + {F\over z_{24}z_{13}}
+ {G\over z_{23}^2}\Big]\,.
\end{equation}
In  this parametrization  $A$ to $F$ are of order  $\oA{2}$ and
$G$ is  of order $\oA{}$. The twelve domains of integration are
given in eq.~(\ref{eOrd}).

\medskip

There  is some  freedom in  which the OPEs leading to the
expression~(\ref{expre}) are performed~\cite{Mafra:2009bz} that
can  give   an  equivalent form  of  the integrand of the
amplitude. Let us define the quantity
\begin{equation}\label{eCJ}
C_{x,y}^z ={1\over (x-z)(z-y)}\,.
\end{equation}
Clearly this function satisfies the Jacoby identity
\begin{equation}
J(x,y,z)=C_{x,y}^z+C_{z,x}^y+C_{y,z}^x=0\,.
\end{equation}
The freedom in parameterizing the  amplitude in~(\ref{expre})
is given by the possibility of having
\begin{equation}\label{eJac}
J(1,2,3)=0\,,\qquad
J(4,2,3)=0\,.
\end{equation}
In  the  amplitude~(\ref{expre})  we  have  made explicit  the
poles $C_{2,3}^1$ and $C_{1,2}^3$ and $C_{3,4}^2$ and
$C_{2,3}^4$.

This freedom corresponds to local monodromy transformations
exchanging the position of neighboring vertex operators. There
are as well global monodromy transformations given by moving
vertex operators from one side of the line to the other side
which are not captured by these local transformations.

The 12  color-ordered five-point  amplitudes are given by
specifying the range  of  integration  over   $z_2$  and  $z_3$
over  the  following domains\footnote{We have  $(n-1)!/2$ such
domains  corresponding to the different $(n-1)!$ color-ordered
amplitudes divided by 2 by reflection.} of integrations
$D_\sigma$
\begin{equation}
\begin{array}{rl}
D_{12345}&= \{0\leq z_2 \leq z_3\leq 1\}\,,\\ \phantom{AAA} D_{13245}&=\{0\leq z_3 \leq z_2\leq 1\}\,, \\
D_{12435}&= \{0\leq z_2 \leq 1\leq z_3\}\,,\\  D_{13425}&= \{0\leq z_3 \leq 1 \leq z_2\}\,,  \\
D_{14235}&= \{0\leq 1\leq z_2 \leq z_3\}\,,\\  D_{14325}&= \{0\leq 1\leq z_3 \leq z_2\}\,,  \\
D_{21345}&= \{ z_2 \leq 0\leq z_3\leq 1\}\,,\\  D_{31245}&= \{ z_3 \leq 0 \leq z_2\leq 1\}\,,  \\
D_{23145}&= \{ z_2 \leq z_3\leq 0\}\,,     \\   D_{32145}&= \{ z_3 \leq z_2\leq 0\}\,,  \\
D_{21435}&= \{ z_2\leq 0\leq 1\leq z_3\}\,,\\  D_{31425}&= \{ z_3\leq 0 \leq 1\leq z_2\}\,.
\end{array}
\label{eOrd}
\end{equation}
We now use the result for $I(a,b,c,d,e)$ which is given in the
appendix A. The integrals are explicitly evaluated in
appendix~\ref{sec:5pointintegrals}. We here quote the field
theory results. In the field theory limit $\ap\to0$ we get
\begin{align}
{\cal A}_5(1,2,3,4,5) ={}& \frac{A}{s_{12}s_{45}} +
\frac{B-Gs_{34}}{s_{23}s_{51}}+\frac{C}{s_{34}s_{12}}+
\frac{E+Gs_{13}}{s_{45}s_{23}}+\frac{D-Gs_{34}}{s_{51}s_{34}}\,,
\label{stringone}
\end{align}
\begin{align}
{\cal A}_5(1,3,4,2,5) =&\frac{A-E-F}{s_{13}s_{25}} - \frac{D-Gs_{34}}{s_{34}s_{51}} +
\frac{-F}{s_{42}s_{13}}
-\frac{D-C}{s_{25}s_{34}}+\frac{B-D}{s_{51}s_{42}}\,,
\label{stringtwo} \\ \nonumber \\
{\cal A}_5(1,2,4,3,5) =&\frac{A-C}{s_{12}s_{35}}+\frac{B-D}{s_{24}s_{51}}-
\frac{C}{s_{43}s_{12}} +\frac{F+B-D}{s_{35}s_{24}}-\frac{D-Gs_{34}}{s_{51}s_{43}}\,,
\label{stringthree} \\ \nonumber\\
{\cal A}_5(1,3,2,4,5) =&\frac{A-E-Gs_{13}}{s_{13}s_{45}}-\frac{B-Gs_{34}}{s_{32}s_{51}}
-\frac{-F}{s_{24}s_{13}} -\frac{E+Gs_{13}}{s_{45}s_{32}}-\frac{B-D}{s_{51}s_{24}}\,,
\label{stringfour} \\ \nonumber \\
{\cal A}_5(1,4,3,2,5) =& \frac{D-C+A-E-F}{s_{14}s_{25}} +
\frac{D-Gs_{34}}{s_{43}s_{51}} + \frac{B-E+Gs_{35}}{s_{32}s_{14}}\nonumber \\
& + \frac{D-C}{s_{25}s_{43}} + \frac{B-Gs_{34}}{s_{51}s_{32}}\,,
\label{stringfive} \\ \nonumber \\
{\cal A}_5(1,4,2,3,5) =& \frac{D-C+A-F-B-Gs_{35}}{s_{14}s_{35}} -
\frac{B-D}{s_{42}s_{51}} - \frac{B-E+Gs_{35}}{s_{23}s_{14}}\nonumber \\
& - \frac{F+B-D}{s_{35}s_{42}} - \frac{B-Gs_{34}}{s_{51}s_{23}}\,.
\label{stringsix}
\end{align}
It is interesting to note that we could use monodromy relations
for integrals on the individual $A$, $B$, $C$ etc. terms in
\eqref{expre}. Thereby one would obtain the same relations as
for the full subamplitudes, but now just for the individual
terms. Hence, the OPEs provide us with expressions for the
subamplitudes in which the relations are very explicitly
reduced to relations in the pole structure. This can also be
checked explicitly for the five-point case by use
of~\eqref{stringone}--\eqref{stringsix}.

\subsection{The generalized parametrization (from strings)}
In~\eqref{stringone}--\eqref{stringsix} we already wrote the
amplitudes in terms of double poles. The quantities $A$ to $F$
were naturally put into the double-pole form, but the $G$ term,
a single-pole term, was forced into this representation by
making a specific choice. Later we will come back to the
freedom in absorbing the $G$ terms, but for now we just
consider the form given above.

Comparing with Bern, Carrasco and
Johansson's~\cite{Bern:2008qj} parametrization ({\it
i.e.}~\eqref{partialfive1}--\eqref{partialfive6}) we identify
from \eqref{stringone}--\eqref{stringsix}
\begin{eqnarray}
\begin{array}{lclcl}
n_1 = A\,, &\phantom{A} & n_6 = D-C+A-E-F\,, &\phantom{A} & n_{11}=B-D\,, \\
n_2= B-Gs_{34}\,, & & n_7= B-E+Gs_{35}\,, & & n_{12}=A-C\,, \\
n_3=C\,, & & n_8=D-C\,, & & n_{13}=F+B-D\,, \\
n_4=E+Gs_{13}\,, & & n_9=A-E-F\,, & & n_{14}=D-C+A-F-B-Gs_{35}\,, \\
n_5 = D-Gs_{34}\,, & & n_{10} = -F\,, & & n_{15} = A-E-Gs_{13}\,.
\end{array}
\end{eqnarray}
The Jacobi-like identities then take the form
\begin{eqnarray}
&&X_1 = n_3-n_5+n_8 =Gs_{34}\,, \nonumber  \\
&&X_2 = n_3-n_1+n_{12} =0\,, \nonumber \\
&&X_3 = n_4-n_1+n_{15}=0\,, \nonumber \\
&&X_4 = n_4-n_2+n_7= -Gs_{32}\,, \nonumber\\
&&X_5 = n_5-n_2+n_{11}=0\,, \nonumber\\
&&X_6 = n_7-n_6+n_{14}=0\,, \nonumber\\
&&X_7 = n_8-n_6+n_9=0\,, \nonumber\\
&&X_8 = n_{10}-n_9+n_{15}=-Gs_{13}\,, \nonumber\\
&&X_9 = n_{10}-n_{11}+n_{13}=0\,.
\end{eqnarray}
And from~\eqref{mone}--\eqref{mfour} it is easy to see that
these amplitudes do indeed satisfy the BCJ-relations. Moreover
not all $X_i$'s vanish.

Note that the BCJ-relations could also be derived from
\eqref{stringone}--\eqref{stringsix} by expressing, for
instance, $A$ and $B$ in terms of two subamplitudes and the $C$
to $G$ terms. Using these expressions
 for $A$ and $B$ in the remaining amplitudes leads directly to BCJ-relations (the $C$ to $G$ terms
 vanish after the substitution).

\subsection{Distributing the single-pole terms}
There are many ways of arranging the $G$ terms into the
numerators of double poles. The expressions given above
correspond to just one specific choice. To see this more
clearly let us begin by defining $\tilde{n}_i$'s
\begin{eqnarray}
\begin{array}{lclcl}
\tilde{n}_1 = A\,, &\phantom{A} & \tilde{n}_6 = D-C+A-E-F\,, &\phantom{A} &
\tilde{n}_{11}=B-D\,, \\
\tilde{n}_2= B\,, & & \tilde{n}_7= B-E\,, & & \tilde{n}_{12}=A-C\,, \\
\tilde{n}_3=C\,, & & \tilde{n}_8=D-C\,, & & \tilde{n}_{13}=F+B-D\,, \\
\tilde{n}_4=E\,, & & \tilde{n}_9=A-E-F\,, & & \tilde{n}_{14}=D-C+A-F-B\,, \\
\tilde{n}_5 = D\,, & & \tilde{n}_{10} = -F\,, & & \tilde{n}_{15} = A-E\,.
\end{array}
\end{eqnarray}
The amplitudes can then, in all generality, be represented like
\begin{align}
{\cal A}_5(1,2,3,4,5) &\equiv \frac{\tilde{n}_1+G g_1}{s_{12}s_{45}} +
\frac{\tilde{n}_2+G g_2}{s_{23}s_{51}} +
\frac{\tilde{n}_3+G g_3}{s_{34}s_{12}} +
\frac{\tilde{n}_4+G g_4}{s_{45}s_{23}} +
\frac{\tilde{n}_5+G g_5}{s_{51}s_{34}}\,,
\end{align}
\begin{align}
{\cal A}_5(1,4,3,2,5) &\equiv \frac{\tilde{n}_6+G g_6}{s_{14}s_{25}} +
\frac{\tilde{n}_5+G g_5}{s_{43}s_{51}} +
\frac{\tilde{n}_7+G g_7}{s_{32}s_{14}} +
\frac{\tilde{n}_8+G g_8}{s_{25}s_{43}} +
\frac{\tilde{n}_2+G g_2}{s_{51}s_{32}}\,, \\ \nonumber \\
{\cal A}_5(1,3,4,2,5) &\equiv \frac{\tilde{n}_9+G g_9}{s_{13}s_{25}} -
\frac{\tilde{n}_5+G g_5}{s_{34}s_{51}} +
\frac{\tilde{n}_{10}+G g_{10}}{s_{42}s_{13}} -
\frac{\tilde{n}_8+G g_8}{s_{25}s_{34}} +
\frac{\tilde{n}_{11}+G g_{11}}{s_{51}s_{42}}\,,\\ \nonumber \\
{\cal A}_5(1,2,4,3,5) &\equiv \frac{\tilde{n}_{12}+G g_{12}}{s_{12}s_{35}} +
\frac{\tilde{n}_{11}+G g_{11}}{s_{24}s_{51}} -
\frac{\tilde{n}_3+G g_3}{s_{43}s_{12}} +
\frac{\tilde{n}_{13}+G g_{13}}{s_{35}s_{24}} -
\frac{\tilde{n}_5+G g_5}{s_{51}s_{43}}\,, \\ \nonumber \\
{\cal A}_5(1,4,2,3,5) &\equiv \frac{\tilde{n}_{14}+G g_{14}}{s_{14}s_{35}} -
\frac{\tilde{n}_{11}+G g_{11}}{s_{42}s_{51}} -
\frac{\tilde{n}_7+G g_7}{s_{23}s_{14}} -
\frac{\tilde{n}_{13}+G g_{13}}{s_{35}s_{42}} -
\frac{\tilde{n}_2+G g_2}{s_{51}s_{23}}\,, \\ \nonumber \\
{\cal A}_5(1,3,2,4,5) &\equiv \frac{\tilde{n}_{15}+G g_{15}}{s_{13}s_{45}} -
\frac{\tilde{n}_2+G g_2}{s_{32}s_{51}} -
\frac{\tilde{n}_{10}+G g_{10}}{s_{24}s_{13}} -
\frac{\tilde{n}_4+G g_4}{s_{45}s_{32}} -
\frac{\tilde{n}_{11}+G g_{11}}{s_{51}s_{24}}\,,
\end{align}
where the $g_i$'s are new parameters representing the fractions
of the $G$ terms absorbed into the specific double poles. Since
these expressions must equal
\eqref{stringone}--\eqref{stringsix} in order to express the
actual amplitudes, we get six equations constraining the $g_i$
parameters
\begin{align}
\frac{s_{13}}{s_{45}s_{23}}-\frac{s_{34}}{s_{23}s_{51}}-
\frac{1}{s_{51}} &= \frac{g_1}{s_{12}s_{45}} + \frac{ g_2}{s_{23}s_{51}} +
\frac{ g_3}{s_{34}s_{12}} + \frac{g_4}{s_{45}s_{23}} + \frac{ g_5}{s_{51}s_{34}}\,,  \\
\frac{s_{35}}{s_{14}s_{23}}-\frac{s_{34}}{s_{23}s_{51}}-\frac{1}{s_{51}}&=
\frac{ g_6}{s_{14}s_{25}} + \frac{ g_5}{s_{43}s_{51}} + \frac{ g_7}{s_{32}s_{14}}
+ \frac{ g_8}{s_{25}s_{43}} + \frac{ g_2}{s_{51}s_{32}}\,, \\
\frac{1}{s_{51}} &= \frac{ g_9}{s_{13}s_{25}} -
\frac{ g_5}{s_{34}s_{51}} + \frac{ g_{10}}{s_{42}s_{13}} -
\frac{ g_8}{s_{25}s_{34}} + \frac{ g_{11}}{s_{51}s_{42}}\,, \\
\frac{1}{s_{51}} &= \frac{ g_{12}}{s_{12}s_{35}} +
\frac{ g_{11}}{s_{24}s_{51}} - \frac{ g_3}{s_{43}s_{12}} +
\frac{ g_{13}}{s_{35}s_{24}} - \frac{ g_5}{s_{51}s_{43}}\,, \\
\frac{s_{34}}{s_{51}s_{23}}-\frac{s_{35}}{s_{23}s_{41}}-
\frac{1}{s_{41}} &= \frac{ g_{14}}{s_{14}s_{35}} -
\frac{g_{11}}{s_{42}s_{51}} - \frac{ g_7}{s_{23}s_{14}} -
\frac{ g_{13}}{s_{35}s_{42}} - \frac{ g_2}{s_{51}s_{23}}\,, \\
\frac{s_{34}}{s_{15}s_{23}}-\frac{s_{13}}{s_{23}s_{45}}-
\frac{1}{s_{45}} &= \frac{ g_{15}}{s_{13}s_{45}} -
\frac{ g_2}{s_{32}s_{51}} - \frac{ g_{10}}{s_{24}s_{13}} -
\frac{ g_4}{s_{45}s_{32}} - \frac{ g_{11}}{s_{51}s_{24}}\,.
\end{align}
Any solution to these equations give a valid distribution of
the $G$ terms, {\it i.e.} provide us with a representation of
the form~\eqref{partialfive1}--\eqref{partialfive6} that
satisfy \eqref{mone}--\eqref{mfour}.

The representation written out explicitly
in~\eqref{stringone}--\eqref{stringsix} corresponds to the
solution
\begin{eqnarray}
\begin{array}{lclcl}
g_1 = 0\,, &\phantom{A} & g_6 = 0\,, &\phantom{A} & g_{11}=0\,, \\
g_2= -s_{34}\,, & & g_7= s_{35}\,, & & g_{12}=0\,, \\
g_3=0\,, & & g_8=0\,, & & g_{13}=0\,, \\
g_4=s_{13}\,, & & g_9=0\,, & & g_{14}=-s_{35}\,, \\
g_5 = -s_{34}\,, & & g_{10} = 0\,, & & g_{15} = -s_{13}\,.
\end{array}
\end{eqnarray}

A numerical check have shown that there \textit{do} exits
solutions for $g_i$ such that the nine Jacobi identities
($n_i-n_j+n_k = 0$) are satisfied, and in such a way that four
of the $g_i$'s can be chosen arbitrarily. This correspond to
the freedom Bern, Carrasco and Johansson find in choosing their
$\alpha_1$, $\alpha_2$, $\alpha_3$ and $\alpha_4$ arbitrarily.

An example of a (valid) choice of $g_i$'s which generate
$n_i$'s that satisfy the Jacobi identities is
\begin{eqnarray}
\begin{array}{lclcl}
g_1 = -s_{12}\,, &\phantom{A} & g_6 = -s_{25}\,, &\phantom{A} & g_{11}=0\,, \\
g_2= -s_{12}-s_{25}\,, & & g_7= -s_{25}\,, & & g_{12}=0\,, \\
g_3=-s_{12}\,, & & g_8=-s_{25}\,, & & g_{13}=0\,, \\
g_4=-s_{12}\,, & & g_9=0\,, & & g_{14}=0\,, \\
g_5 = -s_{12}-s_{25}\,, & & g_{10} = 0\,, & & g_{15} = 0\,,
\end{array}
\end{eqnarray}
with, {\it e.g.}
\begin{align}
n_3-n_5+n_8 &= (\tilde{n}_3-\tilde{n}_5+\tilde{n}_8) + G(g_3-g_5+g_8) \nonumber \\
& = (C-D+D-C) + G(-s_{12}-(-s_{12}-s_{25})-s_{25}) \nonumber \\
& = 0, \qquad etc\ldots
\end{align}
From the expansion given by the OPE this might not be the most
simple or natural way of absorbing the $G$ terms into
double-poles, but it does show that the assumption of Bern,
Carrasco and Johansson is allowed for (at least) the five-point
case.

\section{Monodromy and KLT relations}
As a direct application of the monodromy relations in Yang-Mills theory,
we can rewrite the Kawai-Lewellen-Tye relations at four-point level
in the following manner
\begin{equation}
\cM_4={\kappa_{(4)}^2\over
\alpha'}\,{\cSS{k_1}{k_2}\cSS{k_1}{k_4}\over\cSS{k_1}{k_3}}\,\cA_4^L(1,2,3,4)
\cA_4^R(1,2,3,4)\,.
\label{KLTFour}
\end{equation}
The field theory limit of  the  string
amplitude~(\ref{KLTFour}), $\alpha'\to0$  gives  the symmetric
form of the gravity amplitudes of~\cite{Bern:2008qj}
\begin{equation}
M_4   =\kappa_{(4)}^2\,{st\over   u}\,\left({n_s\over  s}+{n_t\over
t}\right)\left({\tilde n_s\over s}+{\tilde n_t\over
t}\right)
=-\kappa_{(4)}^2\,\left({n_s\tilde n_s\over s}+{n_t\tilde n_t\over
t}+{n_u\tilde n_u\over u}\right) \,.
\label{KLTFourFT}
\end{equation}
Here we have made use of the on-shell relation $s+t+u=0$ and
the four-point Jacobi relation $n_u=n_s-n_t$.

At five point order Bern, Carrasco and
Johansson~\cite{Bern:2008qj} showed that if the subamplitudes
are parameterized by numerators like in
eqs.~\eqref{partialfive1}--\eqref{partialfive6}, and we assume
the numerators satisfy the Jacobi-like identities, then the KLT
relation
\begin{align}
-iM_5(1,2,3,4,5) ={}& s_{12}s_{34}A_5(1,2,3,4,5)\widetilde{A}_5(2,1,4,3,5) \nonumber \\
&\phantom{AAAAAA} + s_{13}s_{24}A_5(1,3,2,4,5)\widetilde{A}_5(3,1,4,2,5)\,,
\end{align}
implies the following form of $M_5$
\begin{align}
-iM_5(1,2,3,4,5) ={}& \frac{n_1\tilde{n}_1}{s_{12}s_{45}} + \frac{n_2\tilde{n}_2}{s_{23}s_{51}}
+ \frac{n_3\tilde{n}_3}{s_{34}s_{12}} + \frac{n_4\tilde{n}_4}{s_{45}s_{23}}
+ \frac{n_5\tilde{n}_5}{s_{51}s_{34}} \nonumber \\
&+\frac{n_6\tilde{n}_6}{s_{14}s_{25}} + \frac{n_7\tilde{n}_7}{s_{32}s_{14}} + \frac{n_8\tilde{n}_8}{s_{25}s_{43}}
+ \frac{n_9\tilde{n}_9}{s_{13}s_{25}} + \frac{n_{10}\tilde{n}_{10}}{s_{42}s_{13}} \nonumber \\
&+ \frac{n_{11}\tilde{n}_{11}}{s_{51}s_{42}}+ \frac{n_{12}\tilde{n}_{12}}{s_{12}s_{35}}
+ \frac{n_{13}\tilde{n}_{13}}{s_{35}s_{24}}+ \frac{n_{14}\tilde{n}_{14}}{s_{14}s_{35}}
+ \frac{n_{15}\tilde{n}_{15}}{s_{13}s_{45}}\,.
\label{grafive}
\end{align}
If we instead use the more general solution for $A_5$ and
$\widetilde{A}_5$, {\it i.e.}
\begin{eqnarray}
\begin{array}{lclcl}
X_1 \equiv s_{34}f_1\,, & & X_2 \equiv s_{12}f_2\,, & & X_3 \equiv s_{45}f_3\,, \\
X_4 \equiv s_{23}f_4\,, & & X_5 \equiv s_{15}f_5\,, & & X_6 = s_{14}(f_1-f_2+f_4)\,, \\
X_7 = s_{25}(f_1-f_3+f_4)\,, & & X_8 = s_{13}(f_3-f_1-f_5)\,, & & X_9 = s_{24}(f_3-f_2-f_5)\,,
\end{array}
\end{eqnarray}
and
\begin{eqnarray}
\begin{array}{lclcl}
\widetilde{X}_1 \equiv s_{34}g_1\,, & & \widetilde{X}_2 \equiv s_{12}g_2\,, & & \widetilde{X}_3 \equiv s_{45}g_3\,, \\
\widetilde{X}_4 \equiv s_{23}g_4\,, & & \widetilde{X}_5 \equiv s_{15}g_5\,, & & \widetilde{X}_6 = s_{14}(g_1-g_2+g_4)\,, \\
\widetilde{X}_7 = s_{25}(g_1-g_3+g_4)\,, & & \widetilde{X}_8 = s_{13}(g_3-g_1-g_5)\,,
& & \widetilde{X}_9 = s_{24}(g_3-g_2-g_5)\,.
\end{array}
\end{eqnarray}
Here $X_1 = n_3'-n_5'+n_8'$ and $\widetilde{X}_1 =
\tilde{n}_3'-\tilde{n}_5'+\tilde{n}_8'$, see eq.~\eqref{Xsol},
and we obtain
\begin{align}
-iM_5(1,2,3,4,5) ={}& \frac{n_1'\tilde{n}_1'}{s_{12}s_{45}} + \frac{n_2'\tilde{n}_2'}{s_{23}s_{51}}
+ \frac{n_3'\tilde{n}_3'}{s_{34}s_{12}} + \frac{n_4'\tilde{n}_4'}{s_{45}s_{23}}
+ \frac{n_5'\tilde{n}_5'}{s_{51}s_{34}} \nonumber \\
&+\frac{n_6'\tilde{n}_6'}{s_{14}s_{25}} + \frac{n_7'\tilde{n}_7'}{s_{32}s_{14}}
+ \frac{n_8'\tilde{n}_8'}{s_{25}s_{43}} + \frac{n_9'\tilde{n}_9'}{s_{13}s_{25}}
+ \frac{n_{10}'\tilde{n}_{10}'}{s_{42}s_{13}} \nonumber \\
&+ \frac{n_{11}'\tilde{n}_{11}'}{s_{51}s_{42}}+ \frac{n_{12}'\tilde{n}_{12}'}{s_{12}s_{35}}
+ \frac{n_{13}'\tilde{n}_{13}'}{s_{35}s_{24}}+ \frac{n_{14}'\tilde{n}_{14}'}{s_{14}s_{35}}
+ \frac{n_{15}'\tilde{n}_{15}'}{s_{13}s_{45}} \nonumber \\
&-\big[ f_1g_1+f_2g_2+f_3g_3+f_4g_4+f_5g_5 \nonumber \\
&\phantom{-\big[}+f_1(g_4-g_3)+g_1(f_4-f_3) \nonumber \\
&\phantom{-\big[}+f_2(g_5-g_4)+g_2(f_5-f_4) \nonumber \\
&\phantom{-\big[}-f_3g_5-g_3f_5\big]\,.
\end{align}

This representation of the gravity is of course guaranteed to be
exact due to the KLT-construction. We obtain the simple factorized form
(\ref{grafive}) only when we choose
\begin{align}
f_1g_1+f_2g_2+f_3g_3+f_4g_4+f_5g_5
+f_1(g_4-g_3)+g_1(f_4-f_3)\phantom{AAAAAA} \nonumber \\
\phantom{AAAAAAA}+f_2(g_5-g_4)+g_2(f_5-f_4)
-f_3g_5-g_3f_5 = 0\,.
\label{gravterm}
\end{align}
This is evidently satisfied when the numerators fulfill the simple
Jacobi-like relations. However, more general
parameterizations are consistent with this equation as well.
For instance, eq.~\eqref{stringone}--\eqref{stringsix} implies
\begin{align}
f_1 = G, \quad f_4 = -G, \quad \mathrm{and} \quad f_2=f_3=f_5=0\,,
\end{align}
and using the same parametrization for $\widetilde{A}_5$,
eq.~\eqref{gravterm} is seen to be satisfied:
\begin{align}
f_1g_1 + f_4g_4 +f_1g_4 + g_1f_4 = G^2 + G^2 - G^2- G^2 = 0\,.
\end{align}	
Again, the freedom in choosing different representations of the
KLT-relations arise from the freedom to pick parameterizations
of the gauge invariant amplitudes in terms of different pole
structures. These pole structures are not gauge invariant by
themselves and we see that this arbitrariness in the gauge
theory is inherited in the gravity amplitude.

\section{One-loop coefficient relations}

We end this paper with an obvious application of the monodromy
relations in the field theory limit. We illustrate how these relations
can imply relations between coefficients of
integrals in one-loop gluon amplitudes. For simplicity we will
focus on amplitudes in $\mathcal{N}=4$ super
Yang-Mills, but it will be evident that most of the considerations here will
apply also to the case of less supersymmetric or
even non-supersymmetric amplitudes.

\subsection{Preliminaries}
Our starting point will be the one-loop gluon amplitudes which can be color
decomposed~\cite{Bern:1990ux} as follows
\begin{equation}
\mathcal{A}_n^{1-loop} = g^n\sum_{c=1}^{[n/2]+1} \sum_{\sigma\in S_n/S_{n;c}}\mathrm{Gr}_{n;c}(\sigma)A_{n;c}(\sigma)\,.
\end{equation}
Here $[x]$ is the largest integer less than or equal to $x$. The leading color factor is
\begin{equation}
\mathrm{Gr}_{n;1}(\sigma) = N_c\mathrm{Tr}(T^{a_{\sigma(1)}}\cdots T^{a_{\sigma(n)}})\,,
\end{equation}
and the subleading color factors ($c>1$) are
\begin{equation}
\mathrm{Gr}_{n;c}(\sigma) = \mathrm{Tr}(T^{a_{\sigma(1)}}\cdots T^{a_{\sigma(c-1)}})
\mathrm{Tr}(T^{a_{\sigma(c)}}\cdots T^{a_{\sigma(n)}})\,.
\end{equation}
$S_n$ here denotes the set of all permutations of $n$ objects. $S_{n;c}$ is the
subset leaving $\mathrm{Gr}_{n;c}$ invariant.

It is sufficient to consider the subamplitude $A_{n;1}$ which is leading in
color counting, since the
remaining $A_{n;c}$ subamplitudes with $c>1$ can be obtained as a sum over
different permutations of $A_{n;1}$~\cite{Bern:1990ux,Bern:1994zx}.

In $\mathcal{N}=4$ super Yang-Mills theory we can always write the
one-loop gluon amplitude (using a Passarino-Veltman
reduction~\cite{Passarino:1978jh}) as a linear combination of scalar
box integrals with rational coefficients~\cite{Bern:1994zx,Bern:1993kr}.
For the leading subamplitude the expression becomes
\begin{equation}
A_{n;1} = \sum \left(\widehat{b}I^{1m} + \widehat{c}I^{2m\, e} + \widehat{d}I^{2m\, h}
+ \widehat{g}I^{3m} + \widehat{f}I^{4m}\right)\,.
\end{equation}
Here the sum runs over color-ordered box diagrams, and the integrals (defined
in dimensional regularization) are given by
\begin{equation}
I = -i(4\pi)^{2-\epsilon} \int \frac{d^{4-2\epsilon}l}{(2\pi)^{4-2\epsilon}}
\frac{1}{l^2(l-K_1)^2(l-K_1-K_2)^2(l+K_4)^2}\,.
\end{equation}
The external momenta $K_i$ are given by the sum of momenta of consecutive external legs,
and all momenta are taken to be outgoing. The labels $1m,\,2m,\,3m$ and $4m$ refer to the number of
``massive'' corners, {\it i.e.} the number of $K_i^2 \neq 0$. This is equivalent to the number of
corners with more than one external gluon. The $2m$ case is separated into adjacent massive
corners $I^{2m\,h}$ ($h$ for hard), and diagonally opposite massive corners
$I^{2m\,e}$ ($e$ for easy).

Since the scalar box integrals are all known
explicitly~\cite{Bern:1993kr}, calculation of one-loop
amplitudes is reduced to finding the coefficients. From that
general setting the existence of relations between coefficients
of different one-loop amplitudes is surprising. The indication
of such structures does not appear until we introduce unitarity
cuts~\cite{Bern:1994zx,Bern:1994cg}. Working in complex momenta
it is possible to do quadruple cuts and derive formulas for
general coefficients~\cite{Britto:2004nc}
\begin{equation}
\widehat{a}_{\alpha} = \frac{1}{2} \sum_{S,J} n_J A_1^{\mathrm{tree}}A_2^{\mathrm{tree}}
A_3^{\mathrm{tree}}A_4^{\mathrm{tree}}\,.
\label{coform}
\end{equation}
Here $\alpha$ represent a specific ordering of external legs,
$J$ the spin of a particle (running in the loop)
in the $\mathcal{N}=4$ multiplet, $n_J$
the number of particles in the multiplet with spin $J$ and $S$
is the set of the two solutions to the on-shell conditions
\begin{equation}
S = \{\,\, l\,\,\, |\,\,\, l^2=0, \quad (l-K_1)^2 = 0, \quad (l-K_1-K_2)^2=0, \quad (l+K_4)^2=0 \,\, \}\,.
\label{2sol}
\end{equation}

It turns out that for many amplitudes eq.~\eqref{coform}
simplifies significantly. The helicity configuration often
kills the sum over non-gluonic states and one of the $S$
solutions. These coefficients are therefore only given by a
single term of four tree-level gluon amplitudes multiplied
together. Monodromy relations on these tree amplitudes then
leads to relations among coefficients for one-loop amplitudes.
Most interesting is probably the possibility of relating
coefficient for split-helicity loop amplitudes to
mixed-helicity loop amplitudes. For some reviews of the work at
tree and loop level involving helicity amplitudes for gluons,
see {\it e.g.}
refs.~\cite{TreeReview,Cachazo:2005ga,Bern:2007dw}.

\subsection{Six-point examples}
In the following section we give two explicit examples of how the
monodromy relations, in combination with unitarity cuts,
can be used to obtain relations between scalar box integral
coefficients of different one-loop amplitudes.
These should be sufficient to get the idea for more general
one-loop amplitudes.

\subsubsection{Two-mass (easy) coefficient relation}
Let us begin by considering the $\widehat{c}_1$ coefficient to
the $A_{6;1}(1^+,2^-,3^-,4^+,5^+,6^+)$ one-loop amplitude, {\it
i.e.} the coefficient to the $I^{2m\, e}$ integral for a
specific ordering of the legs. Here we choose the one
illustrated in fig. \ref{cut1}. Note that with this helicity
configuration fig. \ref{cut1} is the only diagram that
contributes to $\widehat{c}_1$. Any other assignment of
helicities to the loop-legs makes at least one of the corners
vanish. In addition, only gluons can run in the loop for this
helicity configuration -- fermions and scalars would make the
two corners with equal helicity vanish.

\FIGURE[h]{
\centering
\includegraphics[width=8.7cm]{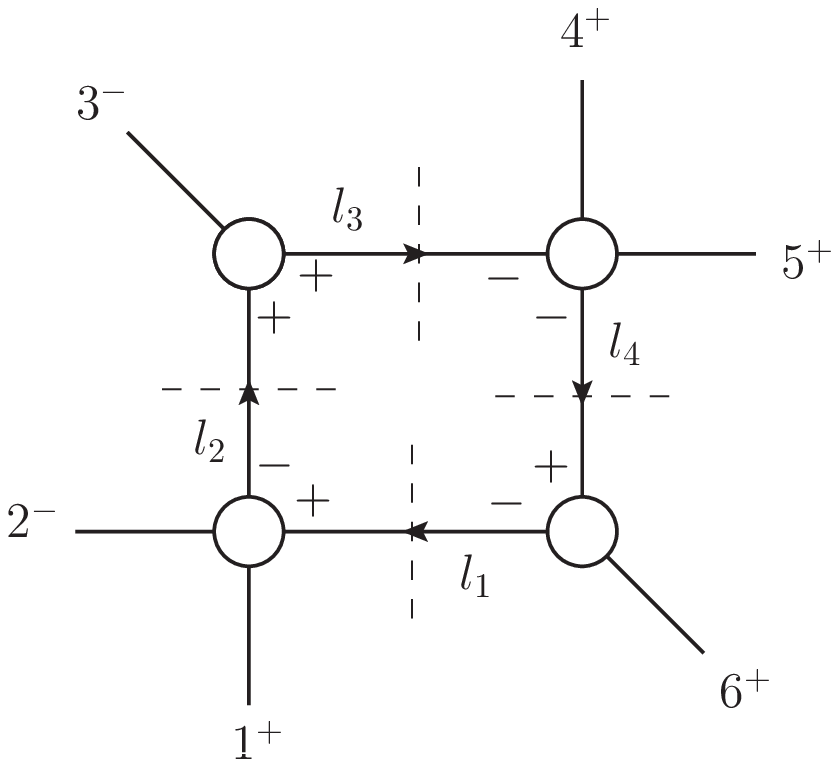}
\caption{Two-mass (easy) cut diagram.}
\label{cut1}
}

Since the four corners are just given by the appropriate
(on-shell) tree-level amplitudes,
we can use the four-point monodromy relations to flip the legs around.
One of the advantages of the monodromy relations is that we can always
keep two of the legs fixed. This is important here since we do not want
to change the position of legs in the loop.
The diagram in fig. \ref{cut1}, which we
denote $\mathcal{D}^{2m\,e}_{12}$,
is therefore related to the diagram of same type,
but with legs 1 and 2 interchanged, through
\begin{equation}
\mathcal{D}^{2m\,e}_{21} = \frac{s_{(-l_1)1}}{s_{l_21}} \mathcal{D}^{2m\,e}_{12}\,.
\end{equation}
The helicity configuration $(++-)$ of the two three-point
corners is only consistent with one of the $S$ solutions
\cite{Britto:2004nc}, and the coefficient is simply given by
$\widehat{c}_1 =\mathcal{D}^{2m\,e}_{12} / 2$. The same is of
course true in the case of leg 1 and 2 interchanged, which
imply that
\begin{equation}
\widehat{c}_1 = \frac{s_{(-l_1)1}}{s_{l_21}} \widehat{c}_1'\,,
\end{equation}
where $\widehat{c}_1'$ is the coefficient to the $I^{2m\, e}$ scalar box integral for the
one-loop amplitude
$A_{6;1}(2^-,1^+,3^-,4^+,5^+,6^+)$. This is a very simple relation between
coefficients for split-helicity and
mixed-helicity loop amplitudes.

For completeness,
we show how to solve for the loop-momenta and express the fraction in
front of $\widehat{c}_1'$ solely in terms of external momenta.
For this we will be using the spinor
helicity formalism. From momentum conservation and
on-shell conditions we have
\begin{equation}
\begin{array}{ll}
l_2 = l_1 - p_1-p_2, &\phantom{AAA} (l_1-p_1-p_2)^2=0, \\
l_3 = l_2 - p_3 = l_1 -p_1-p_2-p_3, &\phantom{AAA} (l_1-p_1-p_2-p_3)^2=0, \\
l_4 = l_3-p_4-p_5 = l_1+p_6, &\phantom{AAA} (l_1+p_6)^2=0\,,
\end{array}
\end{equation}
and in terms of spinor products
\begin{equation}
\frac{s_{(-l_1)1}}{s_{l_21}} = \frac{s_{(-l_1)1}}{s_{(-l_1)2}}
= \frac{\langle 1l_1\rangle \lbrack l_11\rbrack}{\langle 2l_1\rangle \lbrack l_12\rbrack}\,.
\label{sfactor}
\end{equation}
Since the three-point corners have helicity configuration $(++-)$ we
must take the holomorphic spinors
at these corners to be proportional
and hence having vanishing $\langle \bullet \rangle$ product
(remember, we are working with complex momenta,
so the $\lbrack \bullet \rbrack$ product can be non-vanishing).
In particular we get
\begin{equation}
\langle l_1 6\rangle = 0 \quad \Longrightarrow
\quad |l_1\rangle = \alpha |6\rangle\,.
\end{equation}
The proportionality factor $\alpha$ can be obtained from
\begin{equation}
(l_1-p_1-p_2)^2 = 0 \quad \Longrightarrow \quad 2l_2\cdot (p_1+p_2) = (p_1+p_2)^2\,,
\end{equation}
and since $2l_2\cdot (p_1+p_2) = \langle l_1|1+2|l_1\rbrack = \alpha \langle 6|1+2|l_1\rbrack$,
\begin{equation}
\alpha = \frac{(p_1+p_2)^2}{\langle 6|1+2|l_1\rbrack}\,.
\end{equation}
To express the anti-holomorphic spinor of $l_1$ we use
\begin{equation}
(l_1 - (p_1+p_2+p_3))^2 = 0 \quad \Longrightarrow \quad 2l_1\cdot (p_1+p_2+p_3) = (p_1+p_2+p_3)^2\,,
\end{equation}
and
\begin{equation}
 2l_1\cdot (p_1+p_2+p_3) = \langle l_1|1+2+3|l_1\rbrack = \alpha \langle 6|1+2+3|l_1\rbrack\,,
\end{equation}
from which follows
\begin{eqnarray}
&& (p_1+p_2)^2 \langle 6|1+2+3|l_1\rbrack = \langle 6|1+2|l_1\rbrack (p_1+p_2+p_3)^2 \quad
\Longleftrightarrow \nonumber \\
&& \big[ \underbrace{ (p_1+p_2)^2 \langle 6|(1+2+3) -
(p_1+p_2+p_3)^2\langle 6|(1+2)}_{\equiv \lbrack \gamma |} \big] |l_1\rbrack = 0\,,
\end{eqnarray}
{\it i.e.} $|l_1\rbrack = \beta |\gamma\rbrack$. We are not interested in
the proportionality factor $\beta$ since it cancels out
from eq.~\eqref{sfactor} anyway. Using these expressions for
the spinors of $l_1$, we get, after a bit of rewriting,
\begin{equation}
\frac{s_{(-l_1)1}}{s_{l_21}}  = - \frac{\langle 16\rangle
\langle 23\rangle}{\langle 26\rangle \langle 13\rangle}\,.
\end{equation}

\subsubsection{One-mass coefficient relation}
Let us now consider a one-mass box integral coefficient. As
in the example above we just use the
$A_{6;1}(1^+,2^-,3^-,4^+,5^+,6^+)$ one-loop amplitude to illustrate the
idea. The diagram is given in fig. \ref{cut2},
which we denote as $\mathcal{D}^{1m}_{612}$. Again this helicity configuration
kills all other diagrams and allow only gluons to run in the loop.

\FIGURE[h]{
\centering
\includegraphics[width=8.7cm]{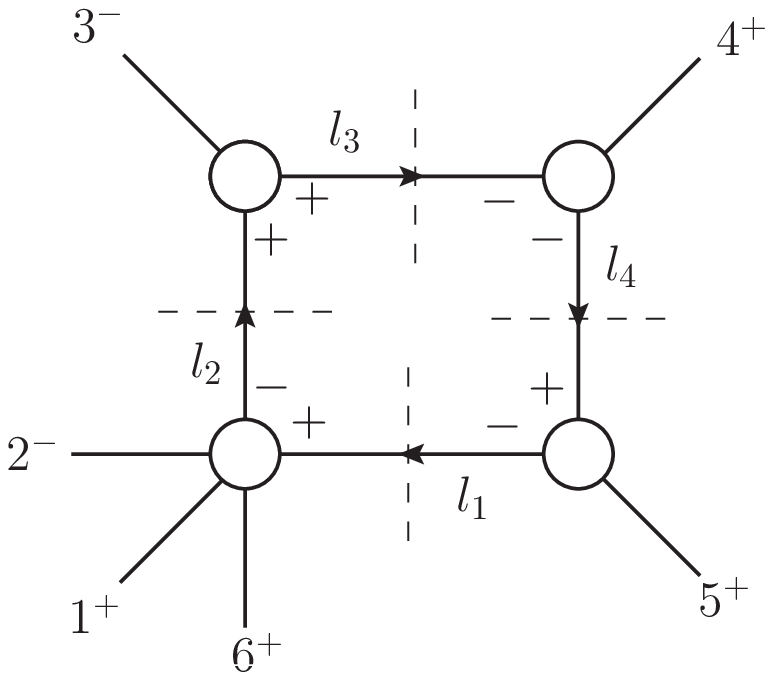}
\caption{One-mass cut diagram.}
\label{cut2}
}

This time we can use the five-point monodromy relations to connect a
diagram of mixed helicity to
two diagrams of split helicities
\begin{equation}
\mathcal{D}^{1m}_{621} = \frac{(s_{16}+s_{(-l_1)1})\mathcal{D}^{1m}_{612}
+ s_{(-l_1)1}\mathcal{D}^{1m}_{162}}{s_{l_21}}\,,
\end{equation}
with obvious notation for the different diagrams. Like above,
the coefficients related to these
diagrams only consist of these single terms, and we can therefore
equally well write it as
\begin{equation}
\widehat{b}_{621} = \frac{(s_{16}+s_{(-l_1)1})\widehat{b}_{612}
+ s_{(-l_1)1}\widehat{b}_{162}}{s_{l_21}}\,,
\end{equation}
where we have a one-mass integral coefficient belonging to the
mixed-helicity amplitude
$A_{6;1}(2^-,1^+,3^-,4^+,5^+,6^+)$ related to
one-mass coefficients of the split-helicity
amplitudes $A_{6;1}(1^+,2^-,3^-,4^+,5^+,6^+)$ and
$A_{6;1}(6^+,2^-,3^-,4^+,5^+,1^+)$.

Using very similar methods as for the two-mass case we could
again express the kinematic invariants in terms of external
momenta. However, this is not our focus here.

\section{Conclusion}

We have reconsidered the BCJ-relations in gauge theories from
several points of view. Based on the monodromy proof, we have
explored the extent to which Jacobi-like relations for residues
of poles (and multiple poles) can be {\em derived}. We have
found that Jacobi-like relations can be introduced consistently
with the constraints of the monodromy relations. But extended
Jacobi-like identities are also perfectly consistent with the
gauge invariant relations. We have demonstrated this explicitly
from both field and string theoretic angles.

We have also considered the implications for gravity
amplitudes. Very symmetric forms follows in a simple manner
through using the KLT-relations together with the link posed by
monodromy in the gauge theory side. This direction appears worthwhile
to pursue in the future.

As an application of monodromy relations, we have explicitly
illustrated how these tree-level relations give rise to
non-trivial identities at loop level. The simplest case is that
of ${\cal N} = 4$ super Yang-Mills theory where relations
between one-loop box functions are directly derivable through
quadruple cut techniques. Similar considerations are valid for
less supersymmetric or non-supersymmetric amplitudes as well,
although in such cases the relations are rather more
complicated. There are thus clearly several interesting
directions for future work that will exploit these relations.

\section{Acknowledgments}
NEJBB and TS would like to acknowledge financial support from
the Danish Council for Independent Research (FNU) and the
L\'eon Rosenfeld Foundation, respectively.

%\appendix
%\section{Appendix}\label{sec:A}{\small

\appendix
\section{Evaluation of the five-point integrals}
\def\s(#1,#2){\hat s_{#1,#2}}
\label{sec:5pointintegrals}
In this  appendix we evaluate the
five point amplitudes~(\ref{expre}) for the  ordering
$(1,2,3,4,5)$. We use the result
\begin{eqnarray}\label{eInt}
I(a,b,c,d,e) &  =&  \int_0^1dz_3   \int_0^{z_3}dz_2\,   z_2^a  (z_3-z_2)^b
(1-z_2)^c (1-z_3)^d z_3^e\\
\nn &=&{\Gamma(a+1) \Gamma(b+1) \Gamma(d+1) \Gamma(a+b+e+2)\over \Gamma
(a+b+2) \Gamma (a+b+d+e+3)}\\
\nn && \times \ _3F_2(a+1,-c,a+b+e+2;a+b+2,a+b+d+e+3;1)\,,
\end{eqnarray}
that expresses the integral in terms of the hypergeometric
function ${}_3F_2$. We introduce the notation
\begin{align}\label{eIfive}
\nn &I_5(a,b,c,d,e)= {\Gamma(\ap\,s_{12}+a+1) \Gamma(\ap\,s_{23}+b+1) \Gamma(\ap\,\ap\,s_{34}+d+1) \Gamma(\ap\,s_{45}+a+b+e+2)\over \Gamma
(s_{2,13}+a+b+2) \Gamma (\ap\, s_{4,35}+a+b+d+e+3)}\\
\nn &\times\ _3F_2(\ap\,s_{12}\!+\!a\!+\!1,-s_{24}\!-\!c,\ap\,s_{45}\!+\!a\!+\!b\!+\!e\!+\!2;\ap\,s_{2,13}\!+\!a\!+\!b\!+\!2,\ap\,s_{4,35}\!+\!a\!+\!b\!+\!d\!+\!e\!+\!3;1)\,,\\
\end{align}
Setting $\hat{s}_{i,j}=\ap\,s_{i,}$ we have

\noindent{\bf Contribution~A}

The integral is
\begin{eqnarray}\label{eIA}
\nn &&I_5(-1,0,0,0,-1)={1\over \s(1,2)\s(1,5)}\,
\frac{ \Gamma(\s(1,2)+1)\Gamma(\s(1,5)+1)\, \Gamma(\s(2,3)+1) \Gamma (\s(3,4)+1) }
{\Gamma(\s(1,2)+\s(2,3)+1) \Gamma(\s(1,2)+\s(1,3)+\s(2,3)+\s(3,4)+1)}
\\
\nn &&\times
\F32(\s(1,2),-\s(2,4),\s(1,2)+\s(1,3)+\s(2,3);\s(1,2)+\s(2,3)+1,\s(1,2)+\s(1,3)+\s(2,3)+\s(3,4)+1;1)\,,\\
\end{eqnarray}

\noindent{\bf Contribution~B}
\begin{eqnarray}\label{eIB}
\nn &&I_5(0,-1,-1,0,0)={1\over \s(2,3)\s(3,4)}\\
\nn &&\frac{\Gamma (\s(1,2)+1) \Gamma (\s(2,3)+1) \Gamma (\s(3,4)+1)
\Gamma (\s(4,5)+1)}{\Gamma (\s(1,2)+\s(2,3)+1) \Gamma
   (\s(3,4)+\s(4,5)+1)}\\
\nn &&\Big[
\F32(\s(1,2)+1,-\s(2,4),\s(4,5)+1;\s(1,2)+\s(2,3)+1,\s(3,4)+\s(4,5)+1;1)\\
\nn &&-\frac{\s(2,3) (\s(4,5)+1) \,
   }{(\s(1,2)+\s(2,3)+1)
   (\s(3,4)+\s(4,5)+1)}
\F32(\s(1,2)\!+\!1,1\!-\!\s(2,4),\s(4,5)\!+\!2;\s(1,2)
   \!+\!\s(2,3)\!+\!2,\s(3,4)\!+\!\s(4,5)\!+\!2;1)\Big]\,,\\
\end{eqnarray}

\noindent{\bf Contribution~C}
\begin{eqnarray}\label{eIC}
\nn &&I_5(-1,0,0,-1,0)={1\over \s(3,4)}\, \frac{\Gamma (\s(1,2)+2)
\Gamma (\s(2,3)+1) \Gamma (\s(3,4)+1
    ) \Gamma (\s(1,2)+\s(1,3)+\s(2,3)+3)}{\Gamma
   (\s(1,2)+\s(2,3)+3) \Gamma (\s(1,2)+\s(1,3)
    +\s(2,3)+\s(3,4)+3)}\\
\nn &&\times \F32(-\s(2,4),\s(1,2)+2,\s(1,2)+\s(1,3)
    +\s(2,3)+3;\s(1,2)+\s(2,3)+3,\s(1,2)+\s(1,3)
    +\s(2,3)+\s(3,4)+3;1)\,,\\
\end{eqnarray}

\noindent{\bf Contribution~D}
\begin{eqnarray}\label{eID}
\nn&&I_5(0,0,-1,-1,0)={1\over \s(3,4)}\\
\nn&&\times\frac{\Gamma (\s(1,2)+1) \Gamma (\s(2,3)+1) \Gamma (\s(3,4)
    +1) \Gamma (\s(1,2)+\s(1,3)+\s(2,3)+2)}{\Gamma
   (\s(1,2)+\s(2,3)+2) \Gamma (\s(1,2)+\s(1,3)  +\s(2,3)+\s(3,4)+2)}\\
\nn&&\times \, \F32(\s(1,2)+1,\s(1,2)+\s(1,3)+\s(2,3)  +2,1-\s(2,4);\s(1,2)
+\s(2,3)+2,\s(1,2)+\s(1,3)  +\s(2,3)+\s(3,4)+2;1)\,,\\
\end{eqnarray}

\noindent{\bf Contribution~E}
\begin{eqnarray}\label{eIE}
\nn&&I_5(0,-1,0,0,-1)={1\over \s(2,3)\s(4,5)}\\
\nn&&\frac{\Gamma (\s(1,2)+1) \Gamma (\s(2,3)+1) \Gamma (\s(3,4)
    +1) \Gamma (\s(4,5)+1)}{\Gamma
   (\s(1,2)+\s(2,3)+1) \Gamma (\s(1,2)+\s(1,3)  +\s(2,3)+\s(3,4)+1)}\\
\nn&&\times \, \F32(-\s(2,4),\s(1,2)+1,\s(1,2)+\s(1,3)  +\s(2,3);\s(1,2)+
\s(2,3)+1,\s(1,2)+\s(1,3)  +\s(2,3)+\s(3,4)+1;1)\,,\\
\end{eqnarray}

\noindent{\bf Contribution~F}
\begin{eqnarray}\label{eIF}
\nn&&I_5(0,0,-1,0,-1)=\frac{\Gamma (\s(1,2)+1) \Gamma (\s(2,3)+1) \Gamma (\s(3,4)
    +1) \Gamma (\s(1,2)+\s(1,3)+\s(2,3)+1)}{\Gamma
   (\s(1,2)+\s(2,3)+2) \Gamma (\s(1,2)+\s(1,3)  +\s(2,3)+\s(3,4)+2)}\\
\nn&&\times \, \F32(\s(1,2)+1,\s(1,2)+\s(1,3)+\s(2,3)  +1,1-\s(2,4);\s(1,2)+
\s(2,3)+2,\s(1,2)+\s(1,3)  +\s(2,3)+\s(3,4)+2;1)\,,\\
\end{eqnarray}

\noindent{\bf Contribution~G}
\begin{eqnarray}\label{eIG}
\nn&&I_5(0,-2,0,0,0)={\s(1,2)+\s(2,3)\over (\s(2,3)-1)\s(2,3) \,\s(4,5)}\\
\nn&&\times\frac{\Gamma (\s(1,2)+1) \Gamma (\s(2,3)+1) \Gamma (\s(3,4)
    +1) \Gamma (\s(4,5)+1)}{\Gamma
   (\s(1,2)+\s(2,3)+1) \Gamma (\s(4,5)+\s(3,4)+1)}\\
\nn&&\times \, \F32(-\s(2,4),\s(1,2)+1,\s(1,2)+\s(1,3)+\s(2,3);\s(1,2)+
\s(2,3),\s(1,2)+\s(1,3)+\s(2,3)+\s(3,4)+1;1)\,,
\end{eqnarray}

\paragraph{}

\end{document}